\documentclass[twocolumn,twocolappendix]{aastex63}
\usepackage[percent]{overpic}
\usepackage{amsmath}

\graphicspath{{./}{Figures/}}

\turnoffonetrue

\newcommand{\kms}{\rm km~s\ensuremath{^{-1}\,}}
\newcommand{\msun}{\ensuremath{\rm M_\odot}}
\newcommand{\msunyr}{\ensuremath{\rm M_{\odot}\;{\rm yr}^{-1}}}
\newcommand{\Ha}{\ensuremath{\rm H\alpha}}
\newcommand{\Hb}{\ensuremath{\rm H\beta}}
\newcommand{\lya}{\ensuremath{\rm Ly\alpha}}

\newcommand{\lyg}{\ensuremath{\rm Ly\gamma}}
\newcommand{\zla}{\ensuremath{z_{\rm Ly\alpha}}}
\newcommand{\zlya}{\ensuremath{z_{\rm Ly\alpha}}}

\newcommand{\lyb}{Ly$\beta$}

\newcommand{\cmtwo}{\rm{cm}\ensuremath{^{-2}}}
\newcommand{\nhi}{\ensuremath{N_{\rm HI}}}

\newcommand{\fluxunits}{\ensuremath{\rm erg~s^{-1}~cm^{-2}}} 
\newcommand{\sbunits}{\ensuremath{\rm erg~s^{-1}~cm^{-2}~arcsec^{-2}}} 
\newcommand{\lumunits}{\ensuremath{\rm erg~s^{-1}}}

\newcommand{\zsys}{\ensuremath{z_{\mathrm{sys}}}}
\newcommand{\zneb}{\ensuremath{z_{\mathrm{neb}}}}
\newcommand{\lognhi}{\ensuremath{\log{N_{\mathrm{HI}}}}}
\newcommand{\HI}{\ensuremath{\mathrm{H~I}}}
\newcommand{\CII}{\ensuremath{\mathrm{C~II}}}

\newcommand{\CIV}{\ensuremath{\mathrm{C~IV}}}
\newcommand{\SiII}{\ensuremath{\mathrm{Si~II}}}
\newcommand{\SiIII}{\ensuremath{\mathrm{Si~III}}}
\newcommand{\SiIV}{\ensuremath{\mathrm{Si~IV}}}
\newcommand{\OVI}{\ensuremath{\mathrm{O~VI}}}
\newcommand{\rvir}{\ensuremath{R_{\mathrm{vir}}}}
\newcommand{\Tvir}{\ensuremath{T_{\mathrm{vir}}}}
\newcommand{\kpc}{\ensuremath{\mathrm{kpc}}}

\newcommand{\lognhicmtwo}{\ensuremath{\log{(N_{\mathrm{HI}}/\mathrm{cm^{-2}})}}}
\newcommand{\K}{\ensuremath{\mathrm{K}}}
\newcommand{\bturb}{\ensuremath{v_{\mathrm{turb}}}}
\newcommand{\vturb}{\ensuremath{v_{\mathrm{turb}}}}
\newcommand{\vlos}{\ensuremath{v_{\mathrm{LOS}}}}
\newcommand{\logT}{\ensuremath{\log{T}}}
\newcommand{\logTK}{\ensuremath{\log{(T/\K)}}}
\newcommand{\T}{\ensuremath{\mathrm{T}}}
\newcommand{\Mstar}{\ensuremath{\mathrm{M_*}}}
\newcommand{\Mdyn}{\ensuremath{\mathrm{M_{dyn}}}}
\newcommand{\logM}{\ensuremath{\mathrm{\log{(M/\msun)}}}}
\newcommand{\logMstar}{\ensuremath{\mathrm{\log{M_*}}}}
\newcommand{\logMmsun}{\ensuremath{\mathrm{\log{(M_*/\msun)}}}}
\newcommand{\logMdynmsun}{\ensuremath{\mathrm{\log{(M_{\rm{dyn}}/\msun)}}}}
\newcommand{\Dtran}{\ensuremath{D_{\mathrm{tran}}}}
\newcommand{\rff}{\ensuremath{r_{\mathrm{eff}}}}

\newcommand{\sSFR}{\ensuremath{\mathrm{sSFR}}}
\newcommand{\SFRSD}{\ensuremath{\Sigma_{\mathrm{SFR}}}}
\newcommand{\sfrsdunits}{\ensuremath{\mathrm{\msunyr~\kpc^{-2}}}}

\newcommand{\Av}{\ensuremath{{A_{\mathrm{V}}}}}
\newcommand{\cmthree}{\ensuremath{{\rm{cm}^{-3}}}}

\newcommand{\Myr}{\ensuremath{\rm{Myr}}}

\newcommand{\dex}{\ensuremath{\rm{dex}}}

\shorttitle{InCLOSE II: CGM of $z\sim2.3$ Low-Mass Galaxies}
\shortauthors{Nunez-Cravin et al.}

\begin{document}

\title{KBSS-InCLOSE II: First Detailed Insights on the Inner CGM of Low-Mass $z\sim2.3$ Galaxies}

\correspondingauthor{Evan Haze Nunez-Cravin}
\email{ehnunezcravin@astro.ucla.edu}

\author[0000-0001-5595-757X]{Evan H. Nunez-Cravin}
\altaffiliation{UC Presidents and Cal-Bridge Postdoctoral Fellow,}
\altaffiliation{Carnegie Observatories Visiting Postdoctoral Scholar}
\affil{University of California, Los Angeles, 475 Portola Plaza, Los Angeles, CA 90095, USA}
\affil{The Observatories of the Carnegie Institution for Sciences, 813 Santa Barbara Street, Pasadena, CA, USA}
\affil{California Institute of Technology, 1200 E. California Blvd., MC 249-17, Pasadena, CA 91125, USA}

\author[0000-0002-4834-7260]{Charles C. Steidel}
\affil{California Institute of Technology, 1200 E. California Blvd., MC 249-17, Pasadena, CA 91125, USA}

\author[0000-0002-8459-5413]{Gwen C. Rudie}
\affil{The Observatories of the Carnegie Institution for Sciences, 813 Santa Barbara Street, Pasadena, CA, USA}

\author[0000-0001-6196-5162]{Evan N. Kirby}
\affil{California Institute of Technology, 1200 E. California Blvd., MC 249-17, Pasadena, CA 91125, USA}
\affil{Department of Physics, University of Notre Dame, Notre Dame, IN 46556, USA}

\author[0000-0000-0000-0000]{Charis M. Hall}
\affil{California Institute of Technology, 1200 E. California Blvd., MC 249-17, Pasadena, CA 91125, USA}

\author[0000-0001-5847-7934]{Nikolaus Z.\ Prusinski}
\affil{California Institute of Technology, 1200 E. California Blvd., MC 249-17, Pasadena, CA 91125, USA}

\author[0000-0002-1945-2299]{Zhuyun Zhuang}
\affil{Center for Interdisciplinary Exploration and Research in Astrophysics (CIERA), Northwestern University, 1800 Sherman Avenue, Evanston, IL 60201, USA}
\affil{California Institute of Technology, 1200 E. California Blvd., MC 249-17, Pasadena, CA 91125, USA}

\author[0000-0002-5770-2666]{Yuanze Ding}
\affil{California Institute of Technology, 1200 E. California Blvd., MC 249-17, Pasadena, CA 91125, USA}

\author[0000-0003-3509-4855]{Alice E. Shapley}
\affiliation{Department of Physics \& Astronomy, University of California, Los Angeles, 430 Portola Plaza, Los Angeles, CA 90095, USA}

\author[0000-0002-6967-7322]{Ryan F. Trainor}
\affiliation{Department of Physics and Astronomy, Franklin \& Marshall College, 637 College Avenue, Lancaster, PA 17603, USA}
\affiliation{William H. Miller III Department of Physics and Astronomy, Johns Hopkins University, Baltimore, MD 21218, USA}

\begin{abstract}
We present results from an extension to the Keck Baryonic Structure Survey (KBSS) that focuses on the Inner Circumgalactic Medium (CGM) of QSO Line Of Sight Emitting galaxies at $z\sim2.3$ (InCLOSE)\@. We analyze two low-mass galaxies $\log{(\rm{M_*} / \rm{M_\odot})} \leq 9$ that are within small a projected distance of a QSO $\Dtran \leq 50~\kpc$ ($\Dtran/\rvir \leq 0.75$). One galaxy is detected as a bright Lyman-$\alpha$ Emitter with Keck/KCWI \textnormal{(confirmed with follow-up Keck/MOSFIRE spectra),} and the other as a serendipitous line emitter with Keck/MOSFIRE. Both galaxies have nebular and morphological properties consistent with those of typical low-mass $z\sim2.3$ star-forming galaxies.
Analysis of their CGM absorption as seen with Keck/HIRES spectra of the background QSOs shows no detections of low-ionization \textnormal{metal} absorption (low-ions; e.g., Si II), ubiquitous detections of intermediate- (e.g., C IV) and high-ions (O VI), kinematically complex absorption ($\geq 7$ components per galaxy halo) spread over $|\Delta v|\pm \geq 150~\kms$, \textnormal{and no \textit{unambiguously} unbound gas is detected in either galaxy halo.} We analyze the thermal properties of a subset of CGM components, finding that the majority (6/10) have temperatures consistent with heating from the metagalactic UV background ($\log{(T/\rm{K})_{\mathrm{med}}}=4.0$) while the remainder possess short-lived, intermediate temperature gas that would require additional heating or rapid replenishment ($4.6 \leq \log{(T/\rm{K})} \leq 5.1$), \textnormal{the internal energy of these absorbers are dominated by thermal broadening, their internal (turbulent) velocities are all subsonic, while their motions through the halo are likely supersonic}. These results hint that the $z\sim2.3$ CGM changes with stellar mass in terms of kinematic complexity and unbound gas fraction, while thermal properties remain similar. Additionally, the results corroborate findings that the CGM evolves with redshift in terms of ion detection rate, density, unbound gas fraction, and thermal properties.
Future KBSS-InCLOSE studies will expand these preliminary findings by leveraging its growing sample of $z\sim2.3$ galaxy-QSO pairs.
\end{abstract}


\section{Introduction} \label{sec:intro}
Low-mass ($\Mstar < 10^{9}~\msun$) galaxies are much more numerous than more massive $M_* \geq 10^{10}~\msun$ galaxies at all redshifts including $z\sim2.3$ \citep{reddy+2009}, and are predicted to play an important role in ionizing and enriching the circumgalactic medium (CGM) and intergalactic medium (IGM) \citep{wetzel+2015,romano+2023}. 
Further, a substantial fraction of galaxies at $z\sim2.3$ galaxies, colloquially known as ``Cosmic Noon,'' are vigorously star-forming \citep{madau+2014} and driving galaxy-scale outflows \citep{shapley+2011,prusinski+2021}. Based on the observation that local galaxies with $\Mstar<10^9~\msun$ have more $>30\%$ of their total baryonic mass in their CGM (even after accounting for stars, neutral gas, and metals in the ISM) \citep{tremonti+2004,behroozi+2010,mcgaugh+2010}, the CGM of $z\sim2.3$ low-mass galaxies likely contains a significant amount of mass that must be included in our understanding of their evolution throughout cosmic time.

However, it is difficult to observe high redshift low-mass galaxies due to their intrinsic faintness. Specifically, the $R$ band magnitude for a typical $10^{10}~\msun\; z\sim2.3$ star-forming galaxy is $R\sim24$ \citep{steidel+2004}, while low-mass Lyman Alpha Emitting (LAE) galaxies at the same redshift have $R\sim27$ \citep{trainor+2015}. 
As a result, there are few constraints on the physical properties of the CGM of these galaxies (e.g., column density, ionization state, kinematics, and temperature).

Studies at low redshift ($z\sim 0.3$) of the low-mass CGM show a systematic lack of low-ionization metal-enriched gas (low-ions hereafter; e.g., C II),  few detections of intermediate-ions (e.g., \CIV), almost ubiquitous detections of high-ions (e.g., O VI), possession of gas that is mostly bound to the galaxy gravitational potential, all coupled with low \HI\ column densities ($\lognhicmtwo \sim 14$; \citealt{johnson+2017,mishra+2024_CUBS9,zheng+2024,dutta+2025}). Their higher mass counterparts show much higher detection rates of low- and intermediate ions, larger HI column densities, but similar fractions of unbound gas suggesting a mass dependence on CGM properties \citep[e.g.,][]{werk+2014,prochaska+2017,koplitz+2025}.

Recent studies at higher redshift ($z>3$) have found that the CGM of faint LAEs often shows intermediate-ions coupled with low HI column density ($\lognhicmtwo$) and some detections of low-ions (as seen in MgII). The majority of the LAEs analyzed in these samples had impact parameters that very likely exceeded the virial radius ($\rvir$) of the galaxy \citep{muzahid+2020,galbiati+2024,banerjee+2026}. Further, since all of the galaxies were \lya\ selected, (currently) lack rest-optical spectra, and (currently) lack deep multi-band imaging, there are few strong constraints on each of their masses. Therefore it is not yet possible to investigate if there is a mass dependence on CGM properties of the high-$z$ sample yet.

At $z\sim2$, inner CGM studies \textnormal{(within \rvir)} have primarily focused on massive star-forming galaxies ($\Mstar\sim10^{10}~\msun$) where a consensus may be emerging. Structurally, it ($z\sim2$ inner CGM) shows moderate detection rates of low-ions ($>60\%)$, ubiquitous detections of intermediate-ions, and moderate detection rates of high-ions ($50\%$) \citep{rudie+2017,rudie+2019,nunez+2024}. Kinematically, it shows complex metal absorption spread over hundreds of \kms, is composed of at least 10 individual components (up to 31!), and has a high incidence rate of unambiguously unbound gas ($>75\%$) \citep{rudie+2017,rudie+2019,nunez+2024}. Thermally, it often has temperatures consistent with heating from photoionization from the UV background, but frequently ($\sim40\%$) has short-lived, warm gas that would require additional heating and/or rapid replenishment \citep{rudie+2019}. \textnormal{What} remains unclear is if these properties are common for low mass star-forming galaxies.

All of these studies illustrate the need to investigate the mass dependence of the high-$z$ CGM during the crucial missing epoch of $z\sim2-3$.

In the past, efficiently finding high-$z$ low mass galaxies required deep narrowband (NB) images coupled with deep optical and/or NIR spectroscopy. The main limitations of this approach are the observational expense and the narrow redshift ranges that can be searched, i.e., the central wavelength chosen for the NB filter. \citet{trainor+2015} analyzed a sample of more than 300 faint $z\sim2.7$ ($z_{\mathrm{avg}}$ for KBSS QSOs) LAEs in this manner. They found that, compared to more massive $z\sim3$ Lyman Break Galaxies (LBGs), the faint LAEs have significantly higher \lya\ escape fractions ($f_{\mathrm{esc}}\sim30\%$), significantly reduced low-ionization metal covering fraction, low dust content ($\Av < 0.1$), smaller outflow velocities ($v_{\mathrm{FUV, avg}} < 500~\kms$) that are likely larger than their escape velocity ($v_{\mathrm{esc}}\sim 200~ \kms$), and are tracing hot, metal-rich outflows. Most importantly, the faint LAEs are typically characterized by extended \lya\ halos ($r > 30~\kpc$) and low stellar masses ($\Mstar < 10^{9}~\msun$).

This work explicitly showed that targeting LAEs near bright QSOs is a viable and efficient means of studying the low-mass, high redshift CGM because of 1) their ease of detection with modern IFUs on 10-m class telescopes and 2) the benefit of using spectroscopy of the bright background QSO\@ to study the diffuse CGM gas in absorption. The higher redshift $z>3$ studies discussed earlier use this same observing strategy.

The KBSS-InCLOSE survey is an extension of the Keck Baryonic Structure Survey (KBSS) that focuses on the Inner CGM of QSO Line of Sight Emitting galaxies (InCLOSE) at $z\sim2.3$ that is discussed in detail by \citet[][InCLOSE I hereafter]{nunez+2024}. KBSS-InCLOSE is well suited to search for LAES near QSOs because Keck/KCWI is the driving instrument for the survey. KCWI is uniquely equipped to detect faint $z\sim2.3$ LAEs due to its blue sensitivity ($\lambda_{\rm min}=3500$ \AA) that allows for the detection of \lya\ emission down to redshift of $\zlya = 1.9$, which is unique among similar IFUs on other 10-m class telescopes (e.g., VLT/MUSE). 
Further, KBSS footprints are centered on hyper-luminous QSOs (V=16-17) at $z\sim2.5-2.8$ that each have high signal to noise (S/N$\sim50-100$ per resolution element) Keck/HIRES spectra \citep{rudie+2012}. 
Finally, KCWI has been shown to be incredibly efficiency at detecting galaxies at small distances ($\theta\lesssim12''$) to the KBSS QSOs through various QSO subtraction techniques \citep{nunez+2024}.

In this paper, we analyze the first low-mass galaxies in the KBSS-InCLOSE sample. These observations serve as a crucial stepping stone towards probing galaxies of diverse properties in the sample. The galaxies were discovered in the KBSS field Q1623. The Q1623 field includes three bright QSOs: KP77 ($z=2.5353$), \textnormal{KP76 ($z=2.4663$), and KP78 ($z=2.6148$),} each within $\sim3'$ of one another. The two galaxies are within \textnormal{small projected angular distance of $\theta <6''$} from KP77 and KP78, respectively. 
We analyze new KCWI datacubes, new MOSFIRE spectra, and extant ground- and space-based images to measure the nebular and morphological properties of the galaxies. \textnormal{The inclusion of Keck/MOSFIRE rest-optical spectra is unique at high-$z$ and allow for independent corroboration of \zlya that is precise enough to allow for meaningful kinematic comparison with CGM gas.} We then use extant Keck/HIRES spectra to analyze the kinematic, abundance, and thermal properties of their CGM absorption. Lastly and most importantly, we compare the emission properties of the galaxies with the absorption properties of the CGM to gain insights into the galaxy-scale baryon cycle of high redshift low-mass galaxies.

This paper is structured as follows. In Section \ref{sec:data_serendipity} we summarize the new observations, data reduction, data processing, and identification of new objects. In Section \ref{sec:ISM} we analyze the galaxy properties including nebular emission, morphology, and stellar populations. In Section \ref{sec:CGM} we analyze the CGM absorption of each galaxy including common observables such as column density and Doppler width, then we extract the thermal properties of the gas including temperature and turbulent velocity. In Section \ref{sec:discussion} we discuss insights on the evolution of the $z\sim2.3$ CGM with mass by comparing the results with relevant studies. In Section \ref{sec:conclusions} summarizes our findings.

Throughout the paper we adopt a \textnormal{$\Lambda$CDM cosmology} of $H_0=70~\kms\rm{Mpc}^{-1}$, $\Omega_m=0.7$, $\Omega_\Lambda=0.3$. 

\section{KCWI as a Discovery Machine for Low Mass Galaxies} \label{sec:data_serendipity}
The data presented in this paper are part of the KBSS-InCLOSE survey. In summary, KBSS-InCLOSE uses Keck/KCWI and Keck/MOSFIRE observations to identify and confirm star-forming galaxies between redshifts $z=1.9-2.6$ that are at small impact parameters or projected/transverse distances (\Dtran) from a KBSS QSO sightlines $\Dtran \lesssim \rvir \sim80-90~\kpc$ \citep[][]{trainor+2012}. The survey includes existing KBSS ground- and space-based images from Keck/LRIS, Keck/MOSFIRE, and \textit{HST}/WFC3-IR (at minimum) which are used to analyze galaxy morphology and stellar populations, and KBSS Keck/HIRES QSO spectra to analyze the CGM absorption associated with the newly identified galaxies.

We used Keck/KCWI to acquire rest-FUV data cubes with each QSO centered in the pointings. We observed Q1623-KP77 on the nights of 2021 July 6, 2021 September 5, and 2024 September 2 for a total integration time of 2.36 hr with full 360$^\circ$ azimuthal coverage surrounding the QSO out to an angular radius of 12'' (96 kpc at $z=2.3$). KP76 was observed on the nights of 2022 May 28, 2022 May 29, and 2024 September 04 for a total integration time of 1.20 hr with the same azimuthal coverage and surrounding footprint. Conditions were good on all nights with typical seeing ranging between 0.48-0.91''. Our primary configuration used the Medium slicer and BL grating to optimize spatial resolution, spectral resolution, and field of view (FoV). The medium slicer has a FoV of $16.5"\times20.3"$ or $\sim135\times166$ physical kpc (pkpc) at z$\sim$2.3 per pointing. The data are binned 2x2 and have a spatial resolution of 0.69'' and spatial sample of 0.3'' \citep{morrissey+2018}. The BL grating covers the spectral range 3500-5500~\AA\ with spectral resolution of R$\sim$1800 (2.5~\AA\ or 166~\kms resolution, sampled at 1~\AA). 

For Q1623-KP77, we combined our data with the KBSS-KCWI survey data \citep{chen+2021} which had a total exposure time of 3.6 hours centered at a position $7.5"$ S of the QSO such that the QSO falls on the N (top) of the the final mosaic. The region of full overlap between the two surveys had a minimum total exposure time of 1.38 hours (center and north of the QSO), while the maximum total integration region (south of the QSO) is much deeper, with total integration time of 5.96 hours.

We reduced and stacked the KCWI data using the procedures described by \citet{nunez+2024}. In summary, the data were reduced using a custom version of the publicly available KCWI DRP\footnote{\href{https://github.com/Keck-DataReductionPipelines/KCWI\_DRP}{https://github.com/Keck-DataReductionPipelines/KCWI\_DRP}}, after which each reduced cube was combined into a final mosaic using the custom post-DRP pipeline KCWIKit\footnote{\href{https://github.com/yuguangchen1/KcwiKit.git}{https://github.com/yuguangchen1/KcwiKit.git}} (\citealt{kcwikit}; implementation described by \citealt{chen+2021} and \citealt{prusinski+2025}). The final cubes have a spatial sampling of $0.3'' \times 0.3''$ and spectral sampling of 1~\AA/pixel.

\begin{table*}[tb]
\centering
\caption{Observations Summary} \label{tab_observations}
\resizebox{\textwidth}{!}{%
    \begin{tabular}{cccccccc}
    \hline
    \hline
    Instrument     &Object     &RA             &Dec            &$t_{\rm exp}$       &Date            &PI         &P/ID\\
    (Config.)      &-          &(J2000.0)      &(J2000.0)      &(hr)                &(YYYY/MM/DD)    &-          &-\\
    \hline
    \hline
    KCWI           &           &               &               &                    &               &           &\\
    Med/BL         &Q1623      &16:25:48.83    &+26:46:58.80   &5.96                &2021/07/06     &Steidel    &C300\\
                   &           &               &               &                    &2021/09/05     &Steidel    &C249\\
                   &           &               &               &                    &2024/09/02     &Steidel    &C355\\
    Med/BL         &Q1623-KP76 &16:25:48.11    &+26:44:32.96   &1.20                &2022/05/28,29  &Steidel    &C263\\
                   &           &               &               &                    &2024/09/02,04  &Steidel    &C355\\
    \hline
    MOSFIRE        &           &               &               &                    &               &           &\\
    H,K            &HU1        &16:25:48.65    &+26:46:55.77   &3.5,2.4$^a$         &2022/08/17     &Steidel    &C205\\
    H              &           &               &               &                    &2023/09/18     &Steidel    &C409\\
    H              &           &               &               &                    &2024/03/30     &Steidel    &C381\\
    H              &           &               &               &                    &2024/04/25     &           &\\
    H,K            &           &               &               &                    &2024/05/25     &           &\\
    H,K            &BX426b     &16:25:47.92    &+26:44:27.45   &2.7,1.0$^a$         &2022/08/17     &Steidel    &C205\\
    H              &BX426b     &16:25:47.92    &+26:44:27.45   &                    &2023/09/18     &Steidel    &C409\\
    \hline
    LRIS           &           &               &               &                    &               &           &\\
    U$n$,G,R$_S$   &Q1623      &16:25:53.95    &+26:46:26.70   &1.0,0.4,1.1$^b$     &2022/08/24     &Steidel    &C205\\
    \hline
    \end{tabular}}
    \tablenotetext{a}{H, K$_S$ band exposure time.}  
    
    \tablenotetext{b}{U$n$, G, R$_S$ band exposure time.}
\end{table*}

\subsection{Images and Galaxy Sizes} \label{sec:data_kbss}
We use new and extant ground-based images from Keck/LRIS, P200/WIRC, Magellan/Fourstar, and space-based images from \textit{HST}/WFPC2, \textit{HST}-IR/WFC3, to analyze the galaxy's morphology and construct spectral energy distributions (SED).

New ground-based UV and optical images ($U_{\rm n} G {\cal R}$; see, e.g., \citealt{steidel+2003}) were obtained on 2022 August 28, from \textit{Keck}/LRIS \citep{Oke+1995,steidel+2004} as described in Table \ref{tab_observations}. Ground-based H images \citep[published by][]{steidel+2014,strom+2017} were taken using Magellan/FourStar \citep{persson+2013}. Ground-based JK$_S$ images were taken using P200/WIRC \citep{steidel+2004,wilson+2003}. Finally, space-based NIR images were taken with \textit{HST}/WFC3-IR F140W (Erb PID\#12471 2012March17) and \textit{HST}/WFC3-IR F160W (Law PID\#11694 2010August06). Archival, reduced, and science-ready \textit{HST} images were pulled from the Hubble Legacy Archive \citep[HLA][]{HLA_lindsay+2010}. Q1623-KP77 had coverage in \textit{HST}/WFPC2 F450W (Beckwith PID\#8085 1999May16), \textit{HST}/WFPC2 F702W (Steidel PID\#6557 1997May30), \textit{HST}/WFPC2 F814W (Beckwith PID\#8085 1999May16). KP76 was only covered in the footprint of the \textit{HST}/WFC-IR F160W image. DOI: 10.17909/cxz3-7w80. \textnormal{The data have been collated into a single DOI that can be accessed at \dataset[doi:10.17909/cxz3-7w80]{https://doi.org/10.17909/cxz3-7w80}}.

Both galaxies have small projected sizes so we used the \textit{HST}/WFC-IR F160W to leverage its high spatial resolution to measure their sizes. The images are sampled at $0.08''/$ pixel with a resolution of 0.18'' which is 1.48 kpc in diameter or 0.74 kpc radius.
We fit the galaxies with a two-dimensional Gaussian profile to measure their Full Width at Half Maximum (FWHM). We calculate half of the FWHM and set this as the effective radius.

\subsection{QSO Subtraction on Datacubes and Images} \label{sec:data_qsosub}
The brightness of the hyperluminous QSOs can significantly affect the datacubes and images in waus that can artificially change the physical quantities derived from the data. This is especially true for faint galaxies at small projected radii $b\lesssim3-5''$ ($b\lesssim 40~\kpc$ at $z\sim2.3$). We thoroughly discussed QSO removal in InCLOSE I, but briefly summarize the main points here.

For the IFU data we used two methods of QSO subtraction. One method uses the CubePSFSub routine within the package CubExtractor \citep[CubEx herefter; ][]{cantalupo+2019} that removes QSO emission by empirically constructing a pseudo-narrowband PSF from the cube itself. A PSF is constructed for each each wavelength slice that is fitered using a Gaussian profile by 1 pixel in the spatial and spectral direction, has a spectral width of $\pm175 \AA$, and whose normalization is set by the mean flux in the center 2 spaxels of the QSO center, i.e., a running mean filter.
To reduce over-subtraction we mask wavelength layers that include \lya. After subtraction we use the CubeBKGSub package in CubEx, to remove all continuum sources. We refer to these cubes as ``QSO+continuum subtracted'' and use them to search for and extract \lya\ emitters.

The second method uses a spectral approach to subtract the QSOs using a modified version of IFSFIT \citep{rupke_ifsfit_2014,rupke+2017}. The program assumes that each spaxel can be modeled as a linear combination of the QSO spectrum and additional light, e.g., a foreground galaxy. Our implementation of the program has four steps: 1) QSO continuum extraction (from the datacube), 2) QSO subtraction, 3) QSO \lya\ Halo extraction (from the datacube), and 4) QSO \lya\ Halo subtraction. We refer to these cubes as ``QSO-spectrally subtracted'' and use them to recover continuum emitting galaxies, measure their emission/absorption line properties, and corroborate the LAEs detected in the QSO+Continuum subtracted cubes.

We described in detail how QSOs were removed from the ground-/space-based images in \citet{nunez+2024}. In summary, for each imaging band, we constructed an effective point spread function \citep[ePSF][]{anderson+2000,anderson+2016} using non-saturated field stars of comparable brightness to the QSO with the EPSFBUILDER in the PHOTUTILS software package \citep{bradley_2022}.

\subsection{Keck/MOSFIRE} \label{sec:data_instruments}
All NIR spectra used in this work were obtained using Keck/MOSFIRE \citep{mclean+2012}. Observations were conducted using MOSFIRE's configurable slit unit (CSU) to form multi-slit masks using the same approach described in previous KBSS work \citep{steidel+2014,strom+2017,strom+2018,nunez+2024}. The integration times and program IDs are summarized in Table \ref{tab_observations}; each object had a minimum integration time of $\sim 1$ hour.

The field surrounding Q1623 contains many galaxies of interest that were included as part of a multi-slit mask. The H and K band observations used slightly different CSU mask configurations and sky position angles of the instrument FOV such that the \mbox{0.7'' $\times$ 15.0''} slits would include multiple KBSS galaxies (see Section \ref{sec:ISM_morphology_stellarpop}). The galaxies in this work share the same slit as other galaxies. Total integration times of 3.5, and 1.0 hours were obtained in H and K bands, respectively, over the course of the nights of 2022 September 14-16 under good but variable seeing conditions (0.44''-0.60''~ FWHM)\@.

The data were reduced using the publicly-available MOSFIRE DRP\footnote{\href{https://keck-datareductionpipelines.github.io/MosfireDRP/}{https://keck-datareductionpipelines.github.io/MosfireDRP/}}, which produces background-subtracted, flat-fielded, wavelength-calibrated, telluric-absorption corrected, heliocentric-velocity shifted, rectified, and stacked 2D spectrograms for each slit on the CSU mask. We refer the reader to \citet{steidel+2014} for more details on the data acquisition and reduction, and to \citet{strom+2017} for details on the flux calibration.

As described by \citet{nunez+2024}, we use MOSPEC  \citep{strom+2017}\footnote{\href{https://github.com/allisonstrom/mospec}{https://github.com/allisonstrom/mospec}} to extract the 1D spectra from the 2D spectrograms, measure redshifts, linewidths, and fluxes. For each galaxy, we used either a boxcar spatial profile or Gaussian profile to extract the 1D spectra for each band. Slit loss corrections were determined separately for the H and K band spectra using a method described by \citet{strom+2017}. Briefly, the spectrum of a bright calibration star (included on the same mask as the galaxies) was extracted and integrated to calculate the total slit flux. We compared the slit flux to the flux measured from photometry of the same star then adopted the ratio between the slit flux and the photometric flux as the slit loss correction. The slit loss correction for H-band is 1.84 and for K$_S$ band is 1.69.

\subsection{Photometry, Dynamical Masses, \& Star Formation Rate} \label{sec:sed_fit}
We measure the photometry of each galaxy using PHOTUTILS. We use circular apertures with diameters equal to two times \rff\ centered on the galaxy, and two nearby (within few arcseconds) background apertures of the same size to estimate sky background.

Photometry was corrected for bright nebular emission using KCWI or MOSFIRE spectra, in the corresponding band. We corrected 
F160W for [O~III]$\lambda\lambda4960,5008$. The corrections were very small $m_{\rm{emission_correction}}<0.05$.

We calculated the dynamical mass for each galaxy by combining their measured effective radii (F160W) with their line widths measured from non-resonant rest-optical nebular emission  (i.e., [O~III]). We correct the linewidth for the instrumental resolution of MOSFIRE ($\sim30~\kms$ \citet{mclean+2012}). We use the same approach as InCLOSE I and assume that the linewidths measured from the strong lines are proportional to the galaxy's circular velocity. Specifically, $\Mdyn=C\sigma^2 \rff/G$ where $\sigma$ is the measured linewidth, G is the gravitational constant and C a constant which we approximate to $C\sim3.4$ to account for the unknown geometry and projection effects of the galaxy.

We place limits on the star-formation rates for each galaxy using their rest-optical spectra. We convert \Ha\ flux to star-formation rate (SFR) in the same manner as KBSS-InCLOSE I using the conversion factor derived by \citet{theios+2019} to convert $\log{L_{\Ha}}$ to $\rm{SFR_{\Ha}}$. This has the form $\log{SFR}=\log{\Ha}-C$, where C is the conversion factor $C=41.64$. This calibration constant is similar to the constant computed by \citet{clarke+2025} in a recent analysis of high-redshift galaxies best fit with BPASS SPS models and a SMC like extinction curve in the JADES and AURORA JWST surveys. We do not have robust detections of \Ha\ for either galaxy so we use their \Hb\ flux limit to place a limit on \Ha\ by using the intrinsic \Ha/\Hb\ ratio of nebular gas consistent with the high redshift ISM with an electron density of $n_e=100~\cmthree$ and electron temperature of T=15,000 K ($I_{\Ha}/I_{\Hb}=2.79$. This is slightly smaller than the ratio assumed in InCLOSE I. This SFR based on \Hb, which we will refer to as SFR(\Hb) is a lower limit on the true SFR because a dust extinction correction would increase $F_{\Ha}$ leading to a larger SFR.

We calculate the specific star formation rate by normalizing SFR(\Hb) by the dynamical mass. We quote the sSFR in units of ${\rm Gyr^{-1}}$.

We calculate the SFR surface density $\Sigma_{\rm SFR}$ ($\msunyr~\kpc^{-2}$) by combining SFR(\Hb) from MOSFIRE and \rff\ measured from the F160W images. Specifically, we normalize the SFR by the area that the disk occupies $A_{\rm eff}=\pi \rff ^2$.

\subsection{Halo Mass and Virial Properties} \label{sec:data_virial}
To estimate the galaxies' halo masses we use the median fit Stellar-Mass to Halo-Mass relation computed from the UniverseMachine \citep[][]{UniverseMachine_behroozi+2019}. We convert $M_h$ to a virial radius (\rvir) and virial temperature (\Tvir) using the definition described by \citet{bryan+1998} and implemented by Yu Lu\footnote{\href{https://github.com/ylu2010/DarkMatterHaloCalculator?tab=readme-ov-file}{https://github.com/ylu2010/DarkMatterHaloCalculator}}. 

\subsection{Keck/HIRES Spectra and Voigt Profile Fitting} \label{sec:data_hiresvoigtprofile}
We probe CGM absorption associated with the galaxies using high-resolution spectra of the QSOs from Keck/HIRES \citep{vogt+1994}. Q1623-KP77 and Q1623-KP76 were observed as part of the original KBSS program \citep{rudie+2012,steidel+2014,turner+2014}. We refer readers to \citet{rudie+2012} for details on the observations, reduction, continuum normalization, and coaddition of the spectra, which include data obtained from both HIRES and VLT/UVES on UT2 \citep{dekker+2000}. The final spectra have $R \simeq 45,000$ $(\mathrm{FWHM} = 6.7~\kms)$, an average $S/N\sim28-48$ per resolution element, and spectral range $\lambda \sim 3126-10,090~$\AA\  \citep{rudie+2012}. 

We analyze the HIRES spectra by performing Voigt-profile decomposition \textnormal{to measure column density $N$, Doppler width $b$, and redshift $z$ or velocity $v$, }using the same procedures described by \citet{nunez+2024}. \textnormal{This approach has the benefit of not needing to explicitly assume e.g., photoionization equilibrium.} In summary, the HIRES spectra are of high enough quality that we perform the fitting by hand by first interactively making initial guesses on the redshift and column density with VoigtFit \citep{voigtfit}, then we perform the final fit using ALIS \citep{ALIS}. We begin by adding a minimum number of absorption components (usually one component per trough), fit, add components where residuals of $\gtrsim 3\sigma$ are present, refit, check the residuals, then iterate until we achieve a reduced $\chi^2\sim1$. We showed in \citet{nunez+2024} that this manual approach is comparable (and sometimes superior) to more sophisticated approaches using e.g., Bayesian ``cloud-by-cloud'' full photoionization suites \citep[see ][]{nielsen+2022}.

When available we adopted the best-fit \HI\ (\lya) column densities, Doppler widths, and redshifts from \citet[][hereafter R12]{rudie+2012} catalog of more than 6,000 \lya\ absorbers towards the KBSS QSOs. When the \HI\ catalog did not include a system of interest we manually fit the absorption in the same manner described previously. Even though almost all \lya\ components were saturated, we were able to place constraints on each absorber's redshift (velocity) and Doppler width due to the detection of \lyb\ and \lyg\ in the HIRES spectra. Importantly, we did not tie any saturated \HI\ components with metal components due to the inherent ambiguity on the true component structure. We tied a subset of weak (unsaturated) \HI\ components with metal absorbers; these weak unsaturated components preferentially had larger velocities since they were blue or red shifted from the saturated components near $v=0~\kms$ i.e., the systemic redshifts of the galaxy position.

We use the best-fit component structure and best-fit parameters to determine the thermal properties of the CGM gas using the \textnormal{following equation: }
\begin{equation} \label{eq:doppler_expansion}
    b_{\mathrm{d}}^2 = v_{\rm turb}^2 + 2kT/m_{\mathrm{ion}}
\end{equation}
 where $k$ is the Boltzmann constant, $m_{\mathrm{ion}}$ is the mass of the ion being fit, $T$ is the temperature of the gas, \textnormal{and $v_{turb}$ is the non-thermal broadening component, e.g., turbulent velocity} \citep{rudie+2019}. This model explicitly assumes that the gas in each absorber is isothermal and therefore each ion arises from the same gas. \textnormal{To ensure that each absorber's Doppler width was physically reasonable, we set a minimum temperature and minimum turbulent velocity of $\logTK_{\rm min}=4$ and $b_{\mathrm{turb}}=1.0~\kms$. ALIS ensures that the total Doppler width matches the resolution of the spectra.}

We used ALIS to refit each of the absorbers but this time allowed the Doppler width to be separately fit between temperature and turbulent velocity. We only allowed the temperature to vary for components that were tied with ions of different masses. Specifically, we tied \CIV\ with at least one other ion of a different mass with the same kinematic structure e.g., \ion{Si}{4}, \ion{O}{6}. This allowed for unambiguous solutions to Equation \ref{eq:doppler_expansion}. For this round of fitting we started with the best-fit redshift and best-fit column density found from the previous fit, then started at minimum \logT\ ($\logTK = $) and \vturb\ ($\vturb = 1~\kms$). Note that for the components that were not tied to ions of another mass we fix the temperature to $\logTK=4$ and let \bturb\ vary, i.e., using the same fitting procedure as the previous section.

We found during the fitting procedure that some components were so narrow that the data could not distinguish between whether thermal or turbulent broadening dominated the linewidth. This resulted in bi-modal distributions of best fit \logT\ and \vturb\ parameters that heavily depended on the starting input parameters. We investigated this issue in Appendix \ref{sec:apdx_vperrors} by performing Monte Carlo simulations of the best-fit parameters finding that only the narrowest components with $b_d \sim 5~\kms$ showed this issue i.e., the models favored both a small \logT\ and small \vturb. We adopted the median best-fit \T\ and \vturb\ as our parameters and the $\pm 1\sigma$ as the error. 

\textnormal{For saturated components we fit the absorption using the components found from non-saturated ions e.g., fitting \SiIII\ with \SiIV\ components, and report the total column density as a lower limit. For non-detections, we calculate upper limits on the total column density by converting the equivalent width measured from $\pm1,000~\kms$ from the transition wavelength, then convert it to a column assuming the gas is in the linear portion of the curve of growth with the simplified expression from \citep{ellison+2004}; a similar approach to calculate upper limits was used by \citep{rudie+2019}.}

\subsection{New Galaxies Towards Q1623} \label{sec:new_objects}
\begin{figure*}
    \centering
    \includegraphics[scale=0.75]{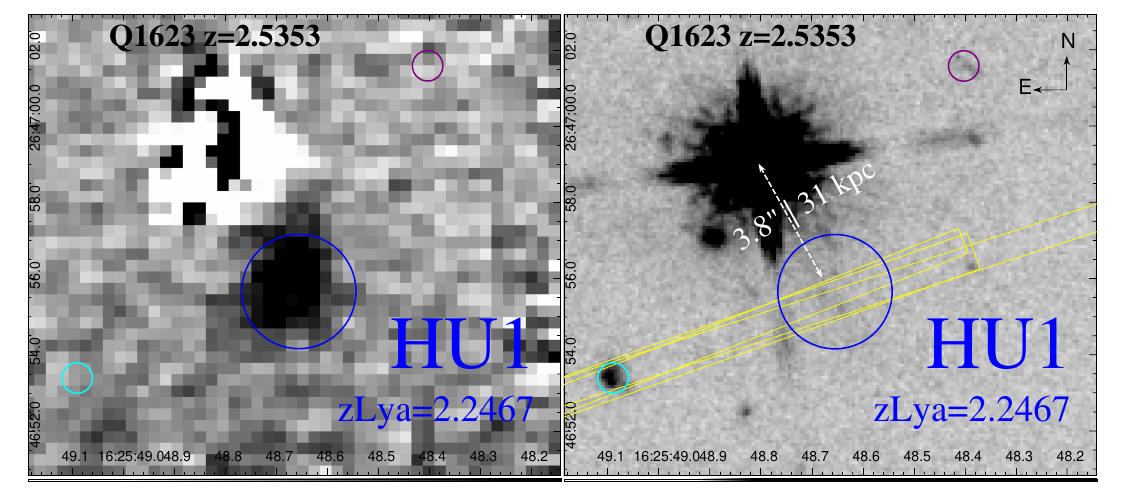}
    \includegraphics[scale=0.75]{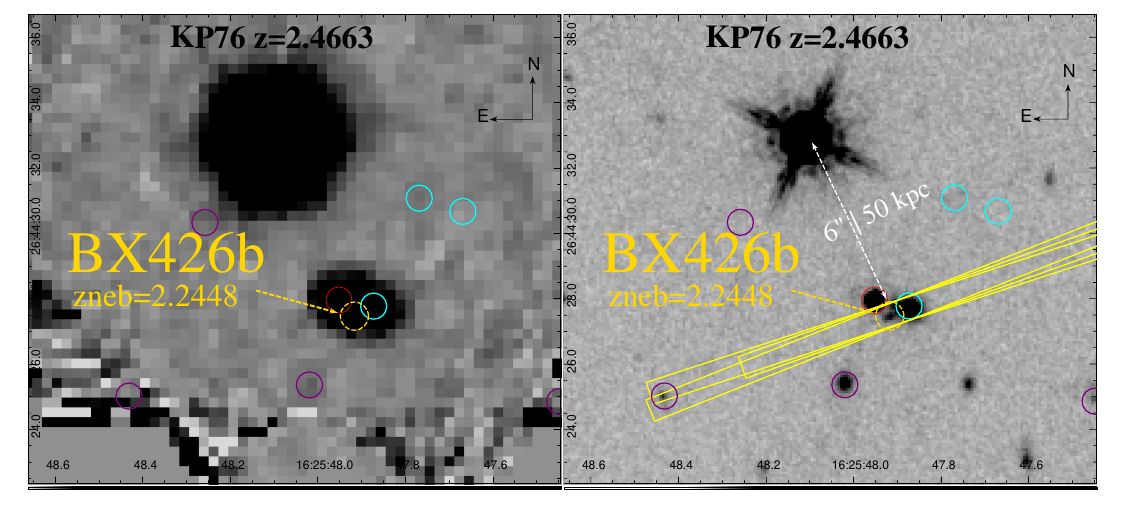}
    \caption{Image of the two galaxies, two QSOs, and newly discovered objects in the Q1623 field as seen with Keck/KCWI (left panels) and \textit{HST}/WFC3-IR F160W (right panels). North is up and east is left. The two QSOs have a projected distance between them of 
    $D(\mathrm {KP77-KP76})=143''=1.17~\rm{Mpc}$. The colored circle positions and sizes are the same for each KCWI (left) and \textit{HST} (right) pair of images. \textit{Top Panels:} Zoom-in on QSO Q1623-KP77 showing galaxy \textnormal{HU1} (blue circle) at a projected distance of $D_{\mathrm{Tran}} = \rm 31~\kpc$. The \textit{top left} shows a pseudo narrowband image centered on \lya\ from KCWI after QSO subtraction (See section \ref{sec:data_qsosub}). The \textit{top right} panel shows the same field of view as the left but with \textit{HST} F160W. Yellow boxes show slits from the multiple MOSFIRE slit masks that included the galaxy (see Section \ref{sec:data_instruments}). \textit{Bottom Panels:} Zoom-in on QSO KP76 showing galaxy \textnormal{BX426b} (blue circle) at a projected distance of $D_{\mathrm{Tran}} = 50~\kpc$. The \textit{bottom left panel} shows a pseudo broadband image from KCWI (3500\AA - 5500 \AA). We can see that the objects of interest are blended. The \textit{bottom right panel} shows the same field of view as the left but with \textit{HST} F160W. In the \textit{HST} image the objects are easily separable. The galaxy shares the same slit as BX426 (cyan circle) which will be discussed in a future InCLOSE paper. A forgreound star (red circle) is in close projected distance to both galaxies. 
    Interestingly both galaxies are at a similar redshift $z=2.24$. 
    }
    \label{fig:main_image}
\end{figure*}

Figure \ref{fig:main_image} shows the two new galaxies analyzed in this paper (blue circles), galaxies that will be discussed in future work (cyan circles), and new galaxies with redshifts of $z<1.9$, $z>z_{\mathrm{QSO}}$, or unknown $z$ detected in the KCWI datacubes (purple circles). 

The top panels of Figure \ref{fig:main_image} show galaxy Q1623-HU1 (HU1 hereafter) at a projected distance from Q1623 of $\Dtran=\rm 31~\kpc$ seen in a QSO subtracted pseudo-narrowband KCWI image (3500-5500 \AA) in the left panel and in a \textit{HST}/WFC-IR F160W image. HU1 is a bright LAE that was discovered while visually inspecting the KCWI cube of Q1623-KP77 (left image). HU1 was bright enough that it was detected \textit{without} the need for QSO subtraction. It is therefore interesting that the only continuum object(s) that is (are) spatially coincident with the \lya\ emission is either a faint clumpy/irregular galaxy or three faint distinct objects, which we will refer to as ``Emission Knots.'' The morphology is shown in the top right panel and discussed in more detail in Section \ref{sec:ISM_morphology_stellarpop}. The yellow boxes show MOSFIRE slits that were included on multiple CSU slit masks designed to ensure that each of the three emission knots were included in the slit i.e., we did not want to ``miss'' the host galaxy. Each slit had slightly different coordinates measured from 1) the centroid of the \lya\ emission peak (cyan contours in the image), 2) the centroid position between the three emission knots seen in the \textit{HST}/F140W image, and 3) the centroid of the brightest emission knot. 

The bottom panels in Figure \ref{fig:main_image} shows galaxy Q1623-BX426b (BX426b hereafter) in a pseudo broadband KCWI image (bottom left panel) and \textit{HST}/WFC-IR F160W image. BX426b is at a projected distance of $\Dtran \sim 50~\kpc$ from the sightline towards KP76. BX426b was discovered serendipitously while searching for rest-optical line emission of brighter galaxy BX426 (to be discussed in future work). BX426 and BX426b were not included in the original KBSS survey because contamination from a nearby foreground star affected the SEDs constructed from the ground-based images. BX426 was added to KBSS-InCLOSE after \lya\ absorption was detected in its (blended) continuum with aforementioned foreground star in the new KCWI datacube (left panel). Only upon observing the bright BX426 on the same MOSFIRE slit mask as HU1 (yellow rectangles) were we able to detect the fainter galaxy examined in this work, BX426b.

\section{Ionized ISM and Stellar Population of the Galaxies} \label{sec:ISM}
In this section we analyze the nebular emission and stellar populations of HU1 and BX426b as seen in KCWI datacubes, MOSFIRE spectra, and ground- and space-based \textit{HST} images. 

\subsection{Nebular Emission} \label{sec:ISM_nebular}
\begin{figure}
    \centering
    \includegraphics[width=0.95\linewidth]{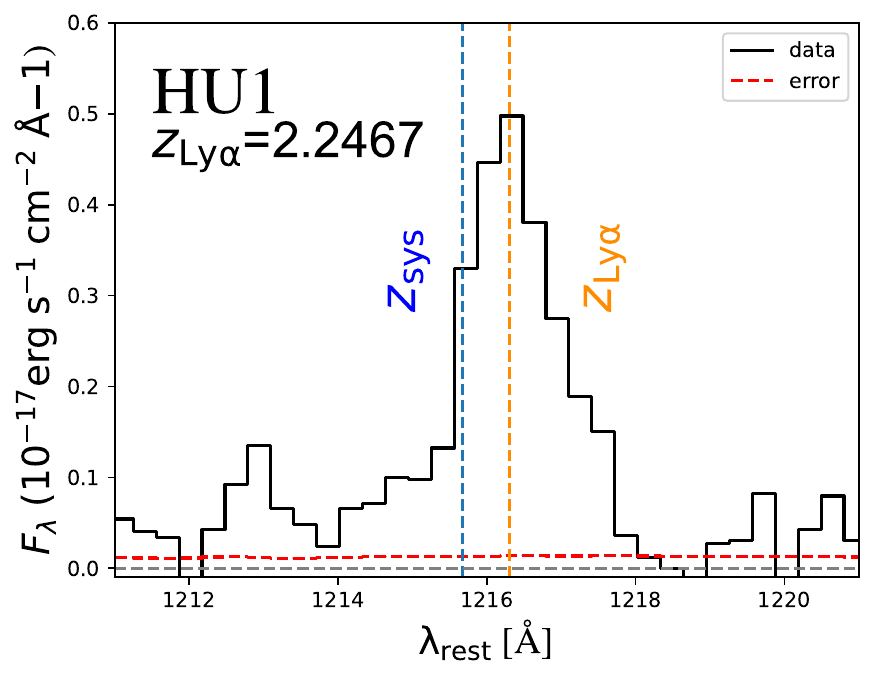}
    \includegraphics[width=1.0\linewidth]{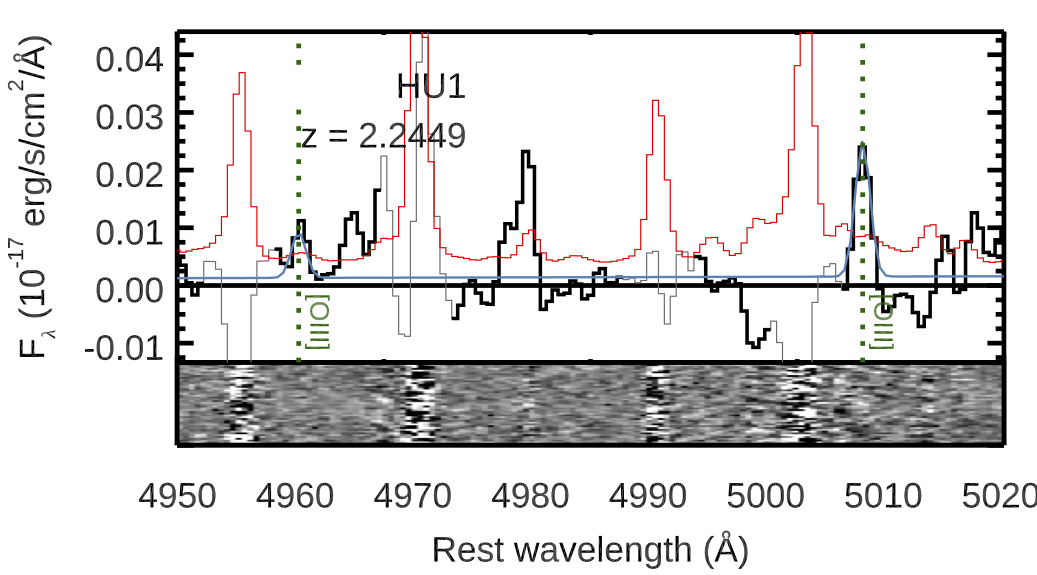}
    \caption{\textit{Top panel:} Extracted KCWI rest-frame FUV spectrum of \textnormal{HU1} from KCWI after QSO subtraction. The dashed blue line shows the systemic redshift of HU1 (measured from [O~III] while the orange line shows the \lya\ redshift. \textit{Bottom Panel:} MOSFIRE H-band spectrum showing a strong detection of $[\mathrm{O~III}]\lambda 5008$. The top of the panel shows the extracted 1D spectrum (black), offset error spectrum (red), and modeled emission lines (green dashed). The bottom shows the 2D spectrogram. We can see two negative images in the 2D spectrogram characteristic of the ABAB mask nod that we observed the mask in with a separation of 1.5'' the main position.
    }
    \label{fig:spec_HU1}
\end{figure}

\begin{figure}
    \centering
    \includegraphics[width=1.0\linewidth]{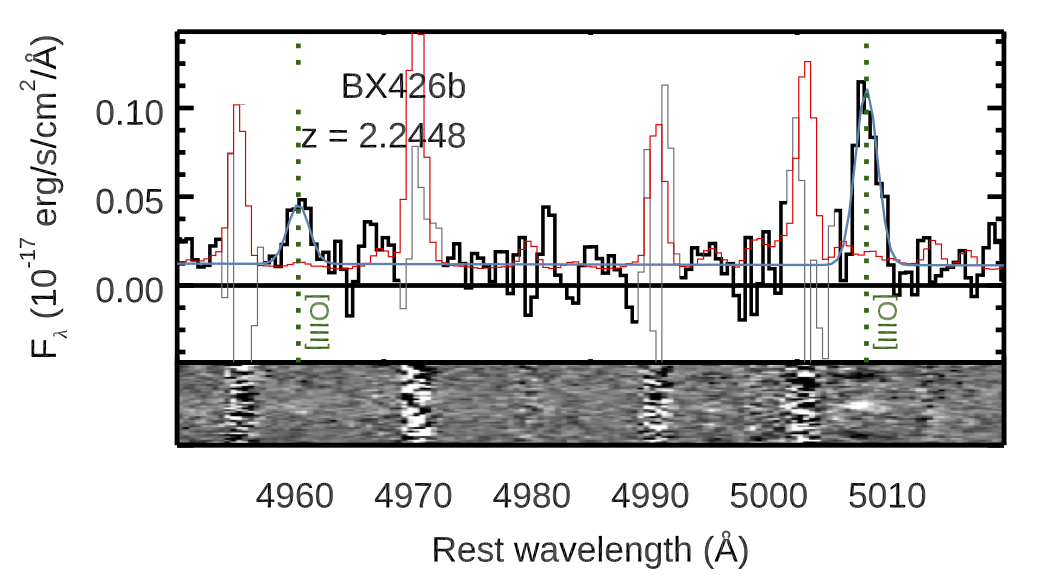}
    \caption{Rest-Optical spectrum of \textnormal{BX426b} from MOSFIRE with the same colors, symbols, and panels as Figure \ref{fig:spec_HU1}. The 1D H-band spectrum shows clear detections of $\mathrm{[O~III]\lambda\lambda4960,5008}$. The 2D spectrogram shows clear negative images corresponding to the ABAB mask nod.
    }
    \label{fig:spec_BX426b}
\end{figure}

Figures \ref{fig:spec_HU1} and \ref{fig:spec_BX426b} show the rest-FUV and rest-optical spectra of HU1 and BX426b, respectively. The spectrum of each galaxy is typical of $z\sim2.3$ star-forming galaxies in that they show redshifted \lya\ emission in the FUV, and ${\mathrm{[O~III]}}\lambda5008$ in the rest-optical.

The top panel of Figure \ref{fig:spec_HU1} shows HU1's strong \lya\ emission. It is the most luminous \lya\ emitter in KBSS-InCLOSE thus far with $\log{(L_{\lya}/\lumunits)}=41.93 \pm 0.1$ ($F_{\lya}=(20.1 \pm 0.4)\times10^{-18}~\fluxunits$) and moderate equivalent width $W_{\rm{rest}}=20.6\pm0.98~\AA$. In stark contrast is its weak $\mathrm{[O~III]}$ emission which is shown in the bottom of Figure \ref{fig:spec_HU1}. We measure a slit-loss corrected flux (flux hereafter) for the $\rm [O~III]\lambda\lambda4960,5008$ doublet of $F_{\rm [O~III]}=(3.50 \pm 0.70)\times10^{-18}~\fluxunits \; (\rm{S/N=5.1})$. We attempted to extract HU1's FUV continuum from the QSO-spectrally subtracted cube but were unable to recover it. The non-detection was due to a combination of object faintness and residuals from the QSO subtraction.

We measure a nebular redshift for HU1 of $z_{\mathrm{[O~III]}}=2.2449$ which is blue shifted by $\Delta v_{\rm{\lya - neb}}=-160~\kms$ from \zla. There is a marginal detection of \Ha\ at $z_{\Ha}=2.2475$ which is a large velocity offset between $\mathrm{[O~III]}$ of $|\Delta z_{\mathrm{[O~III]} - \Ha}|=231~\kms$. We are unaware of a physical mechanism that can explain this velocity difference considering that the emission should arise from the same gas.
Therefore, we do not include this detection in this analysis and instead discuss it, and other faint line emitters, in Appendix \ref{sec:apdx_mosfireserendips}. We measure a marginal detection of \Hb\ below our significance threshold of $S/N>2.5$ so we adopt 2$\sigma$ as an upper limit for its flux, $F(\Hb) < 1.49 ~\times 10^{-18}\fluxunits$. We adopt the systemic redshift as the nebular redshift ([O~III],\Hb) $\zsys = \zneb = z_{\mathrm{[O~III],\Hb}}$. This redshift is similar to the velocity offset between $z_{\rm{[O~III]}}$ and \zla\ of other bright $z\sim2.3$ LAEs \citep{trainor+2015,erb+2023}, and consistent with the velocity range of the CGM absorption (see Section \ref{sec:CGM}).

\begin{table}[th]
\centering
\caption{FUV and Optical Spectral Properties} \label{tab:spectra}
\resizebox{\columnwidth}{!}{%
    \begin{tabular}{ccc}
    \hline
    \hline
    Galaxy                                  &HU1                        &BX426b\\
    
    \hline
    Line Measurements                       &                           &\\
    $\zlya$                                 &$2.2467~(\pm 75~\kms)$     &\nodata\\
    $W(\lya)_{rest}$                        &$20.6 \pm 1.0~$\AA         &\nodata\\
    $z_{\rm{[O~III]}}$                      &$2.2449~(\pm 30.12~\kms)$  &$2.2448~(\pm 22.75~\kms)$\\
    $\sigma_{\rm{[O~III]}}$                 &$\leq 35~\kms$             &$54.94 \pm 6.91~\kms$\\
    
    \hline
    Flux Measurements$^a$                   &$10^{-18}~\fluxunits$      &$10^{-18}~\fluxunits$\\
    $F(\lya)$                               &$20.1 \pm 0.2$             &\nodata\\
    $F(\rm{[O~III]})$                       &$3.50 \pm 0.70$            &$18.1\pm2.2$\\
    $F(\Hb)$                                &$<1.49$                    &$<4.12$\\

    \hline
    Luminosity Measurements                 &(\lumunits)                &(\lumunits)\\
    $\log{L(\lya)}$                         &$41.93 \pm 0.1$            &\nodata\\
    $\log{L(\Ha)}^b$                        &$<42.0$                    &$<42.6^c$\\
    \hline
    \hline
    \end{tabular}
    }
    \tablenotetext{a}{Slit-loss corrected (SC): SC(H)=1.4, SC(K)=1.69}
    \tablenotetext{b}{Calculated assuming $I_{\Ha}/I_{\Hb}$=2.79, where $n_e=100~\cmthree$ and T=15,000 K, \citep{osterbrock+2006}}
    \tablenotetext{c}{Assuming no dust extinction}
    \tablecomments{This table is published in the machine-readable format}
\end{table}

We attempted to recover BX426b from the KCWI cube by constructing a ``star+galaxy-spectrally subtracted" cube (as opposed to "QSO-spectrally subtracted" cubes). This was done by removing the foreground star and the bright galaxy using a similar procedure discussed in Section \ref{sec:data_qsosub}. We searched the star+galaxy-spectrally subtracted cube for continuum emission from BX426b but were unable to recover it. This is mainly because BX426b's small size ($\sim 0.1''$) and KCWI's spatial resolution ($0.69''$) made it difficult to deblend from the bright foreground star and the bright galaxy. 

Figure \ref{fig:spec_BX426b} shows the rest-optical spectrum of BX426b. There is a clear detection of the ${\mathrm{[O~III]}\lambda\lambda4960,5008}$ doublet at $z_{\mathrm{[O~III]}}=2.2448$ with a measured flux of $F_{\mathrm{[O~III]}}=(18.1 \pm2.2)\times10^{-18}~\fluxunits \; ({\rm S/N=8.2})$. This redshift is different from brighter galaxy BX426 which we measure as $z_{\rm [O~III]}\sim2.1$. We do not detect \Hb\ but place a upper limit on its flux by adopting 2$\sigma$ as an upper limit (i.e., integrating the error spectrum at the centroid position of \Hb\ over the same linewidth as [O~III]) which gives $F_{\Hb}<4.12\times10^{-18}~\fluxunits$. We note a marginal detection of \Ha\ at $z_{\Ha}=2.2430$ but do not include it in our analysis due to its large velocity difference with [O~III] of $|\Delta v|=166~\kms$, and much lower signal to noise of $\rm{SNR_{\Ha}}2.6\sigma$. We show \Ha\ and discuss its possible origins in Appendix \ref{sec:apdx_mosfireserendips}. We adopt $\zsys = \zneb = z_{{\rm [O~III]}}$.

The [O~III] line fluxes of HU1 and BX426b are consistent with faint $z\sim2$ LAEs included in KBSS and CECILIA, with $F\lesssim10\times10^{-18}~\fluxunits$. However, their fluxes are small compared to typical $z\sim2.3$ galaxies with stellar masses $\Mstar\sim 10^{10}~\msun$ included in KBSS and MOSDEF, with $F\gtrsim 30\times10^{-18}~\fluxunits$ \citep{steidel+2014,kriek+2015}. This hints that the galaxies may have small stellar masses.

\subsection{Stellar Populations and Morphology} \label{sec:ISM_morphology_stellarpop}
\begin{figure*}
    \centering
    \includegraphics[width=1.0\linewidth]{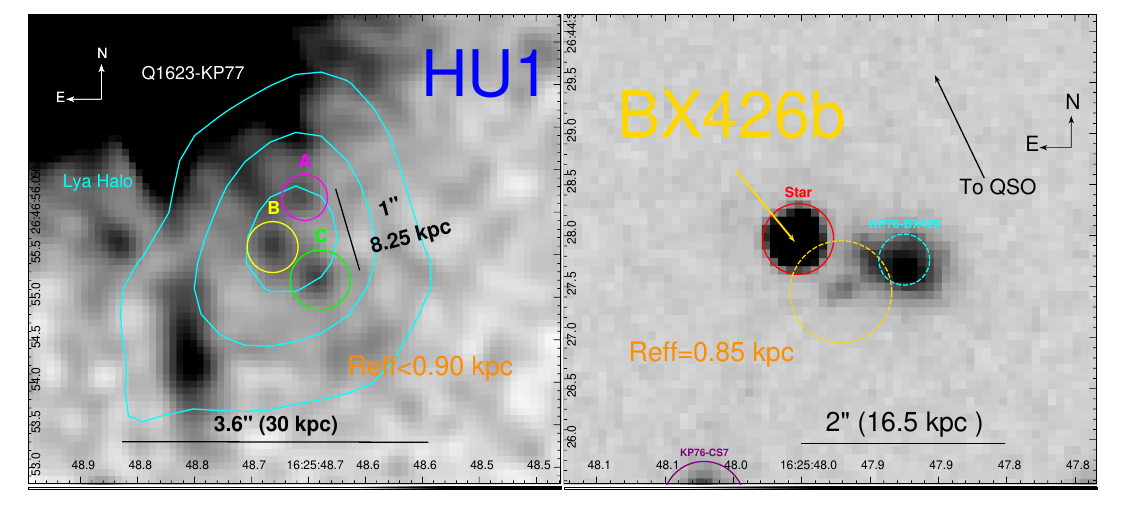}
    \caption{\textit{HST}/F160W rest-optical zoom in images of each galaxy examined in this work. North is up and east is left for all images. The effective radii measured for each galaxy is shown in red. Notably, both galaxies have morphologies and sizes that are reminiscent of low-mass galaxies. \textit{Left:} Zoom-in on \textnormal{HU1} with slight smoothing. Cyan contours show levels of constant \lya\ surface brightness created from a KCWI pseudo-narrowband image at levels 4.8,  3.6,  2.4,  1.2 $ \times$ $10^{-19}$ \sbunits, from the innermost contour to the outermost contour.
    The projected size of the emission extends out to $\sim$30 pkpc and covers all of the emission ``knots'' that may be responsible for the \lya\ emission. The largest distance between the knots (A to C) is $\sim$8 pkpc. \textit{Right panel:} Zoom-in on the morphology of \textnormal{BX426b}. We can see more clearly that it is at a small projected distance from larger galaxy BX426 (dashed cyan circle), and the foreground star (red circle). Though BX426b only occupies $\sim30$ pixels we can see that its morphology is elongated along the slit and narrow perpendicular to the slit suggesting a disk like structure. All three objects are blended together as a single object in ground-based images making it difficult to construct non-model dependent SEDs.}
    \label{fig:pieces}
\end{figure*}

Figure \ref{fig:pieces} showcases the morphology of each galaxy showing that they are well-detected and resolved in the \textit{HST}/F160W images. The right panel of Figure \ref{fig:pieces} shows the morphology of BX426b. The galaxy appears to be extended along the slit in the direction of the bright galaxy (BX426). The average effective radius is $r_{\mathrm{eff, BX426b}}=0.85~\kpc$. We also measure BX426b's magnitude in F160W to be $m_{\rm F160W}=25.3\pm0.3$. \textnormal{This corresponds to an absolute magnitude of $M_{\rm F160W}=-20.9$ and a fraction of $L_*$ in the rest-optical $R$ band of $L/L_* = 0.2$ \citep{marchesini+2007}.} We list these values in Table \ref{tab:photometry}.

\textnormal{We attempted to measure the magnitude and size of each galaxy in the other \textit{HST} images (F450W, F702W, and F814W) but were unable to detect HU1, and BX426b was not included in the footprint of these images. Additionally, we attempted the same measurements with the ground-based images (Keck/LRIS, Magellan/FourStar, P200/WIRC) but were unable to detect HU1, and were unable to separate BX426b from BX426 and the foreground star.}

The left panel of Figure \ref{fig:pieces} shows the morphology of HU1. The cyan contours show the extent of \lya\ overlaid on \textit{HST} rest-optical image. The halo extends to a diameter of 30 \kpc\ (down to surface brightness limit of SB$=1\times10^{-19}~\sbunits$). The centroid of the \lya\ emission is close to the center of three continuum emitting objects, which we will refer to as `` emission knots.'' Each emission knot is faint with magnitudes $m_{\rm F160W}>26.4$, as shown in Table \ref{tab:photometry}. \textnormal{These magnitudes correspond to absolute magnitudes of $M_{\rm F160W}>-19.8$ and in turn $L/L_*<0.07$.} The distance between each knot is shown in the left panel of the Figure where we measure a maximum transverse distance of 8.25 kpc. We measure the effective radius of each knot to be $r_{\mathrm{eff,Knot~A}} < 0.74 ~ \kpc$, $r_{\mathrm{eff,Knot~B}} < 0.74~ \kpc$, and $r_{\mathrm{eff,Knot~C}} = 0.90 \kpc$. Based on the centroid position of \lya, it might be the case that one, or all, of the emission knots is responsible for the emission. We have evidence that either Knot B or Knot C is the host of HU1's \lya\ based on line detections from the different MOSFIRE slitmask configurations. However, we were unable to definitively determine the host due to low signal to noise, which we discuss in Appendix \ref{sec:apdx_emissionknots}.

The measured projected sizes for both galaxies are at largest $\rff=0.90~\kpc$ to $\rff < 0.74~\kpc$ which is consistent with the range of half light radii measured of $z\sim2$ LAEs, which were typically $r_{\mathrm{eff,LAEs}} < 1~\kpc$ \citep{erb+2014}. We tabulate these radii in Table \ref{tab:photometry}. These radii are also consistent with those measured of local dwarf galaxies $r_{\mathrm{eff,Local~Dwarfs}}\sim0.02-2.6~\kpc$ \citep{mcconnachie+2012}.

We calculated the dynamical mass of each galaxy using the measured effective radii and the \textnormal{deconvolved} linewidths from [O~III]$\lambda\lambda4960,5008$ in the previous Section \ref{sec:ISM_nebular}. For HU1 we measure $\sigma_{\mathrm{HU1}}=40.06\pm11.92~\kms$ and for BX426b we measure $\sigma_{\mathrm{BX426b}}=54.94\pm6.91~\kms$. This results in a dynamical mass for BX426b of $\logMdynmsun=9.08\pm0.12$. Since we are not certain which emission ``knot'' is solely responsible for the line emission detected for HU1 (see Figure \ref{fig:pieces}) we calculate \Mdyn\ for each knot finding $\logMdynmsun _{\rm Knot~A,B}<8.34$ and $\logMdynmsun _{\rm Knot~C}=8.43 \pm 0.13$. The dynamical masses of HU1 and BX426b are consistent with the dynamical masses of $z\sim2$ faint LAEs which are typically $\logMdynmsun < 9$ \citep{erb+2014}. 

The dynamical mass is an upper limit of the stellar mass. Since the galaxies in this work are star-forming, and are therefore still building mass, they are not the progenitors of $z\sim0$ dwarf galaxies which have stellar masses defined as $\mathrm{\logMmsun_{Local~Dwarfs}\sim4-9}$ \citep{mcconnachie+2012}. Nonetheless, the dynamical masses confirm that these these galaxies are low-mass since they are 1 dex below M star at this redshift $\logMstar\sim10$ \citep{reddy+2012}.

\begin{table*}[htb]
\centering
\caption{Galaxy Properties} \label{tab:photometry}
\resizebox{\textwidth}{!}{%
    \begin{tabular}{ccccccccc}
    \hline
    \hline
    Galaxy      &$m_{\rm F160W}$    &$L/L_*^a$  &\rff                  &$\log{\Mdyn}$      &SFR$_{\Hb}^a$  &sSFR$_{\Hb}^b$     &$\Sigma_{\rm{SFR}(\Hb)}^a$\\
                &                   &           &($\kpc$)               &($\msun$)          &($\msunyr$)    &($\rm{Gyr^{-1}}$)  &($\msunyr~\kpc^{-2}$)\\
    \hline
    HU1 Knot A  &$>27.6$            &$<0.02$    &$<0.74$                &$\leq8.85$         &\nodata        &\nodata            &\nodata\\
    HU1 Knot B  &$>26.8$            &$<0.05$    &$<0.74$                &$\leq8.85$         &$<0.37$        &$<0.52$            &$<0.21$\\
    HU1 Knot C  &$>26.4$            &$<0.08$    &0.90$^{+0.31}_{-0.23}$ &$\leq 8.94$        &$<0.37$        &$<0.42$            &$<0.14$\\
    BX426b      &$25.3\pm0.3$       &0.21       &0.85$^{+0.27}_{-0.17}$ &$9.08\pm0.12$      &$<1.01$        &$<0.92$            &$<0.44$\\
    \hline
    \hline
    \end{tabular}
}
    \tablenotetext{\textnormal{a}}{\textnormal{$M_{*,R}=-22.67$ in rest-optical $R$ from \citet{marchesini+2007}}}
    \tablenotetext{b}{SFR limits based on \Hb; see Section \ref{sec:sed_fit}}
    \tablecomments{This table is published in the machine-readable format}
\end{table*}

We would like to test whether the galaxies can drive outflows by calculating their star formation rate surface density $\SFRSD \; (\sfrsdunits$) which combines their measured sizes and rest-optical spectra (as described in Section \ref{sec:sed_fit}). \citet{heckman+2015} showed a $\SFRSD > 0.1 ~ \sfrsdunits$ could drive outflows from local galaxies. We have only upper limits on \Hb\ and can only place an upper limit on its SFR \textit{if} we assume that the galaxies have little dust extinction. Given that each galaxy is low mass based on their \Mdyn, small [O~III] line fluxes, and \rff\ similar to that of LAEs, it stands to reason that the dust properties of each galaxy are similar to that of faint $z\sim2$ LAEs, which are characterized by little dust extinction \citep[e.g., $A_{\Ha}\sim0.16$; ][]{trainor+2016}.

With the simplifying assumption of no dust, we estimate both galaxies SFR, sSFR, and \SFRSD and tabulate the results in the fifth, sixth, and seventh columns of Table \ref{tab:photometry}. HU1 (Knot C) has a low SFR of ${\rm SFR(\Hb) < 0.37~\msunyr}$, specific star formation rate of ${\rm sSFR(\Hb) < 0.42~Gyr^{-1}}$, and finally  $\SFRSD({\rm HU1})<0.14 ~ \sfrsdunits$. For BX426b we find $\rm SFR(\Hb)<1.01$, $\rm sSFR(\Hb)<0.92~Gyr^{-1}$, and $\SFRSD({\rm BX426b}) < 0.44~\sfrsdunits$.

The SFR limits on both galaxies are small but the sSFR limits are moderate when compared to the median SFR and sSFR of other $z\sim2.3$ star-forming galaxies analyzed in KBSS \citep{strom+2018}, faint LAEs in KBSS \citep{trainor+2016}, MOSDEF \citep{sanders+2015}, and AURORA \citep{shapley+2025}. The \SFRSD\ all exceed the aforementioned outflow threshold suggesting that both galaxies \textit{could} drive outflows.

Altogether, we showed that the two galaxies in this work: are at similar systemic redshift $\zsys=2.245$ based on their rest-optical nebular emission, are actively star-forming based on \lya\ and/or $[\mathrm{O~III]}$ emission, are weak [O~III] line emitters $F<18\times 10^{-18}~\fluxunits$, have small projected sizes less than $\rff<0.9~\kpc$) based on their effective radii from \textit{HST} images, are low-mass based on their dynamical masses ($\logMdynmsun \leq 9$), are moderately star-forming based on their sSFR, and could be driving galactic outflows based on their \textnormal{star formation rate surface density $\SFRSD \geq 0.15~\sfrsdunits$.}

\section{Inner CGM Properties} \label{sec:CGM}
In this section we infer then analyze the physical properties of the galaxies' CGM based on observables measured in absorption (in the QSO spectra). The galaxies small projected distances from the QSOs ($b<50~\kpc$) and small stellar masses $\mathrm{\log(M_*/M_\odot)<9.0}$ show that we are seeing the first direct high resolution observations of the CGM of low-mass $z\sim2.3$ star-forming galaxies (to our knowledge). Each galaxy has a systemic redshift measured from nebular emission presented in the previous section (Section \ref{sec:ISM_nebular}) which we use to define the zero-point for the line-of-sight (LOS) velocity scale presented henceforth.

\subsection{Column Density and Gas Kinematics} \label{sec:CGM_logN}
\begin{figure*}
    \centering
    \includegraphics[width=0.75\linewidth]{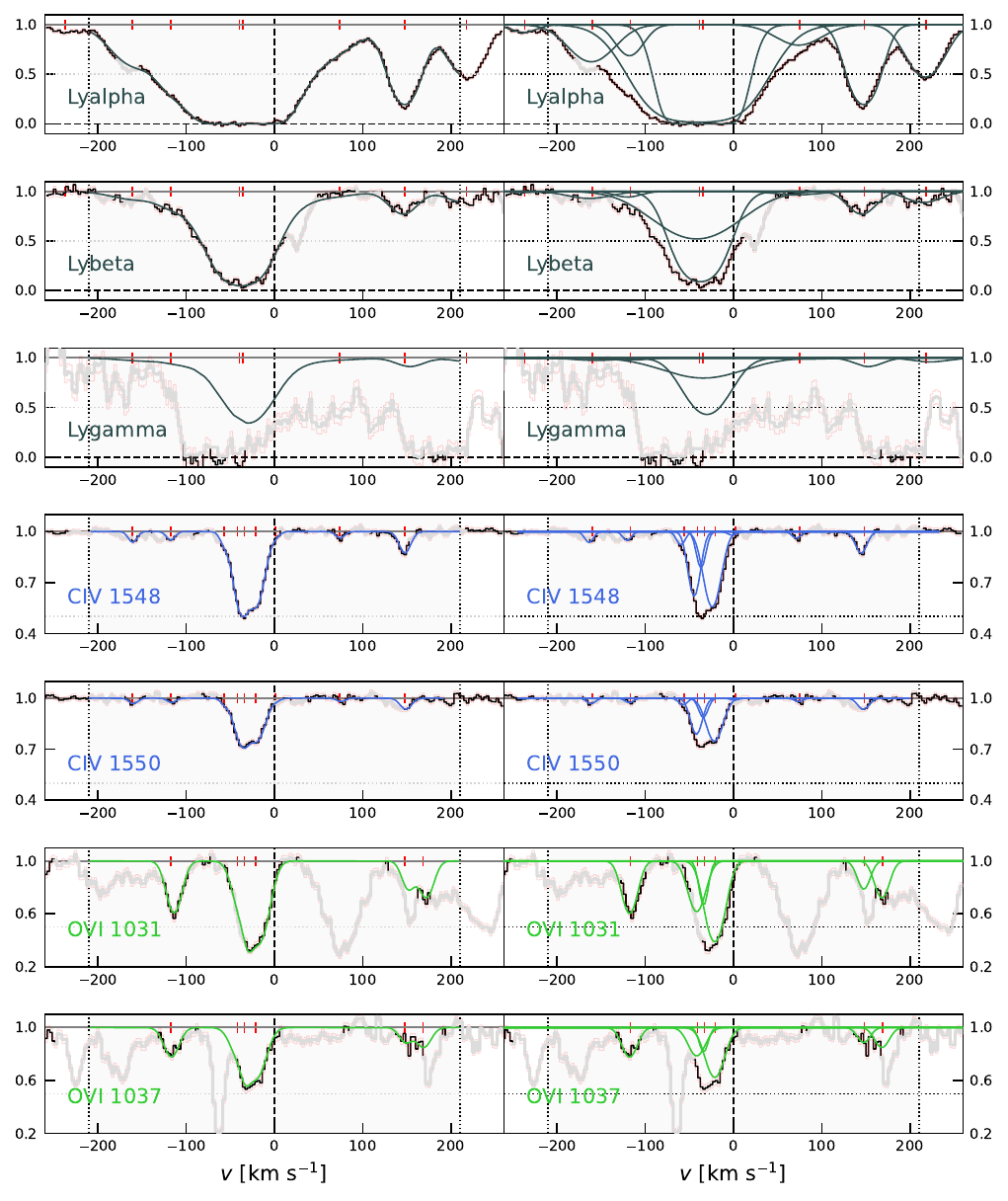}
    \caption{Ion velocity stackplot and best-Fit Voigt-profiles of HU1's CGM absorption centered at  $\zsys=2.2449$. The black lines show the data, the colored lines show the best-fit models, \textnormal{the red curves show the $\pm1\sigma$ error spectrum of the normalized QSO flux,} the dark grey lines show \HI, the light grey lines show intervening contamination that were not included in the fits, \textnormal{the red veritcal markers shows the center of individual Voigt profiles,} and the light grey shaded regions show the extent of the metal-line absorption. \textit{Left:} Product of the individual Voigt profile components, i.e., total fit. \textit{Right:} Individual Voigt-profile components.}
    \label{fig:stackplot_HU1} 
\end{figure*}

\begin{figure*}
    \centering
    \includegraphics[width=0.75\linewidth]{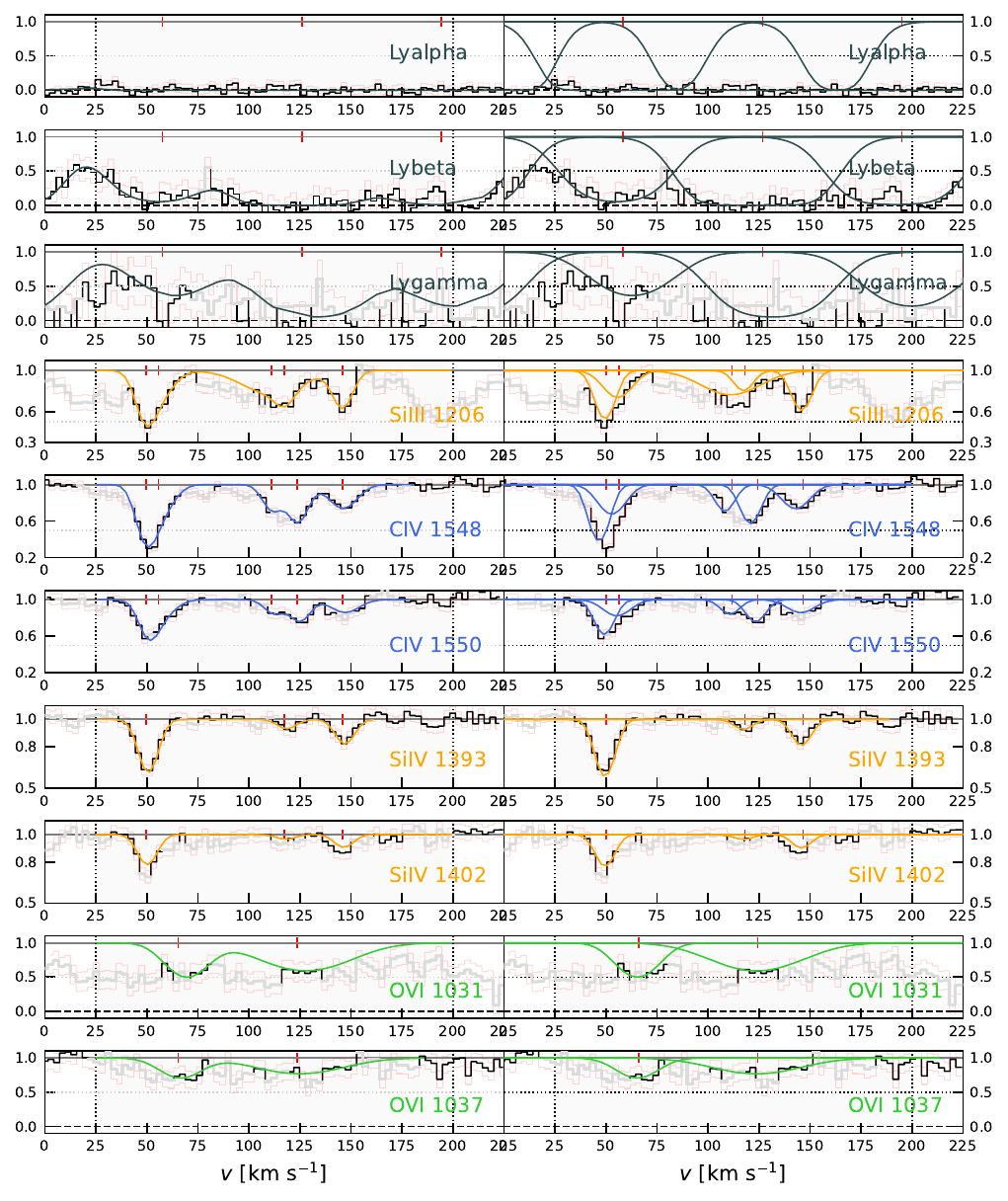}
    \caption{Ion velocity stackplot and best-Fit Voigt-profiles to BX426b's CGM absorption centered at $\zsys=2.2448$. Same colors and lines as Figure \ref{fig:stackplot_HU1}.}
    \label{fig:stackplot_BX426b} 
\end{figure*}

Figures \ref{fig:stackplot_HU1} and \ref{fig:stackplot_BX426b} show ion velocity stackplots with best-fit Voigt profiles overlaid on the data for each galaxy centered at \zsys. Visual inspection of the stackplots shows that each galaxy has a detection of at least \HI, \CIV, \OVI\@. 

Figure \ref{fig:stackplot_HU1} shows HU1's CGM absorption which is spread over a total velocity span of $v\sim400~\kms$. \textnormal{The total \HI\ column density is $\lognhicmtwo=15.04\pm0.18$} which is a couple of dex larger than intergalactic medium absorbers ($\lognhicmtwo \sim 12-14$) but still likely highly ionized \citep{schaye+2003,distefano+2026}. \textnormal{The \CIV\ and \OVI\ total column densities are dominated by two components at $v\sim-20~\kms$ with $\log{(N_{\mathrm{C~IV}}/\cmtwo)}=13.22\pm0.05$, and $\log{(N_{\mathrm{O~VI}}/\cmtwo)}=13.49\pm0.04$.} The strongest \HI\ component is at a similar velocity $v\sim-35~\kms$ with a \textnormal{column density of $\log{(N_{\mathrm{HI}}/\cmtwo)}=14.81\pm0.03$}. 

The kinematics of HU1's CGM absorption are complex in that they are composed of ten unique \textnormal{metal} components spread over $|\Delta v|\sim330~\kms$. The highest velocity components are weak ${\log{(N_{\mathrm{H~I}}/\cmtwo)}}\leq 13.6, \; {\log{(N_{\mathrm{C~IV}}/\cmtwo)}}\leq 12.1, \; {\log{(N_{\mathrm{O~VI}}/\cmtwo)}}\leq 13.4$. We leveraged the weak \HI\ components and kinematically tied all of these weak (non-saturated) together e.g., with \ion{C}{4} and/or \ion{O}{6}. We explore if the components can be modeled as a single phase in the next Section \ref{sec:CGM_Thermal}.

Figure \ref{fig:stackplot_BX426b} shows BX426b's CGM absorption. There is clear \SiIII, \CIV, \SiIV, and \OVI\ spread over $|\Delta v|\sim150~\kms$. The \HI\ is spread over a much larger range of $|\Delta v|\sim350~\kms$. 

We measure a total \HI\ \textnormal{column density of $\lognhicmtwo=15.71\pm0.45$, which is similar to HU1}. We tie \SiIII, \CIV, and \SiIV\ because they show the same kinematic structure. We find a physical, well constrained solution from this component structure. 

The kinematics of BX426b's CGM are moderately complex in that they require seven \textnormal{metal} components (5 unique to \CIV, 1 unique component to \SiIV, and 1 unique to \OVI). The fastest metal ion component \textnormal{(\CIV,\SiIV)} has a velocity of $v\sim150~\kms$ which is at a similar velocity as the \textnormal{strongest \HI\ component $\lognhicmtwo=15.26\pm0.37$.} Interestingly, the strongest metal components \textnormal{(C,Si)} have a velocity near $v\sim50-75~\kms$, which corresponds to a \textnormal{weaker \HI\ component with $\lognhicmtwo=14.75\pm0.09$.} This suggests that the metals are not well-mixed or evenly distributed with \HI. Similar to other absorbers at this redshift, \SiIII, \CIV, and \SiIV\ are kinematically similar in that they are well described by the same component structure. 

The high-ions, as seen in \OVI, have a component structure in the same velocity range as the intermediate-ions, but the components are broader ($b_d =13-30~\kms$) than the intermediate-ions ($b\sim8~\kms$), and only one component shares the same component structure (only one component is tied). In other words, the gas is multiphase which is typical of $z\sim2.3$ CGM absorbers analyzed by \citep{rudie+2019}.

\begin{table*}[htb]
\centering
\caption{CGM Absorption} \label{tab:voigtfits}
\resizebox{1.0\linewidth}{!}{%
    \begin{tabular}{cccccccccc}
    \hline
    \hline
    Galaxy      &QSO    &$z_{\rm sys}$  &$D_{\mathrm{Tran}}$ & & $\log{(\Sigma N_{\rm X})}^a$ & $[(\rm cm^{-2})]$ & &     &$N_{\rm max}^{b}$\\
                &       &               &(pkpc)     &H~I   &Si~III      &Si~IV      &C~IV   &O~VI                           &\\
    \hline
    HU1         &KP77   &2.2449         &31.3       &15.04 &$<$13.7     &$<$12.8    &13.58  &14.23                          &10\\
    BX426b      &KP76   &2.2448         &49.7       &15.71 &$>$12.7     &12.68      &13.62  &14.08                          &7\\
    \hline
    \hline
    \end{tabular}}
    \tablenotetext{a}{The total column density within $\pm1,000~\kms$ of \zsys.}
    \tablenotetext{b}{Maximum number of unique metal components considering all metal ions.}
    \tablecomments{\textnormal{The typical error in the total column density is $<0.45$ dex}.}
    \tablecomments{Lower limits are quoted for saturated components.}
    \tablecomments{\textnormal{Upper limits are quoted for non-detections.}}
    \tablecomments{This table is published in the machine-readable format.}
\end{table*}

Table \ref{tab:voigtfits} summarizes the best fit column densities (summed over $\pm1,000~\kms$) for each ion in the halos of HU1 and BX42b. We place these column densities in context with larger samples in Section \ref{sec:discussion}. We do not include the number of \HI\ components because the they are mostly saturated and likely have unresolved component structure.

\subsection{Temperature and Turbulence} \label{sec:CGM_Thermal}
In this section we investigate the thermal properties of the CGM gas in each galaxy halo by using the best-fit component structure and parameters from the previous Section (see Section \ref{sec:data_hiresvoigtprofile}).

For HU1 we tied \CIV, \OVI, and weak \HI\ components which provide strong constraints on temperature of the gas since the mass ratios are so large (Equation \ref{eq:doppler_expansion}). The fact that we obtain satisfactory and physically reasonable best-fit parameters suggests that these components could be in the same phase. \textnormal{The two components that show the strongest absorption were only tied with \ion{C}{4} and \ion{O}{6} ($v\sim-40~\kms, v\sim-20~\kms$), have widths that are dominated by turbulence ($\bturb = 9.86 \pm 2.72 ~\kms, \bturb = 11.06 \pm 1.24 ~\kms$), with small temperatures ($\logTK = 4.0 \pm 0.37, \logTK = 4.0\pm0.10$); these components were not tied with \HI\ due to \lya\ being saturated. The highest velocity components ($v\sim-160~\kms$, $v\sim140~\kms$) are tied with \HI\ and \ion{C}{4} and are dominated by thermal broadening with the highest temperatures in this sub sample $\logTK = 5.11 \pm 0.22$ and $\logTK = 4.73 \pm 0.02$. The component with the strongest constraints was tied with \HI, \ion{C}{4}, and \OVI\ at $v\sim-120~\kms$ and has both a high temperature $\logTK = 4.60 \pm 0.20$ and large turbulent velocity $\vturb = 9.03 \pm 1.09 ~ \kms$.}

For BX426b's halo we tied \CIV, \SiIV, and \SiIII. Note that we attempted to tie \OVI\ with other ions but were unable to find a physical solution, which showed explicitly that the gas is multiphase. \textnormal{The majority of absorbers in BX426b's halo gas had narrow linewidths ($5.0 \leq b_d \leq  8.6 ~ \kms$) that preferred cool temperature gas $\logTK \sim 4$ \textit{and} low turbulent velocities $\vturb\sim 1-6~\kms$.} \textnormal{The most well-constrained components were tied between \ion{C}{4}, \ion{Si}{3}, and \ion{Si}{4} e.g., the fastest component at $v\sim150~\kms$ had the highest temperature in the halo $\logTK=4.56\pm0.29$ with little turbulent broadening $\vturb=1.00\pm2.03~\kms$.}

\begin{figure*}
    \centering
    \includegraphics[width=0.48\linewidth]{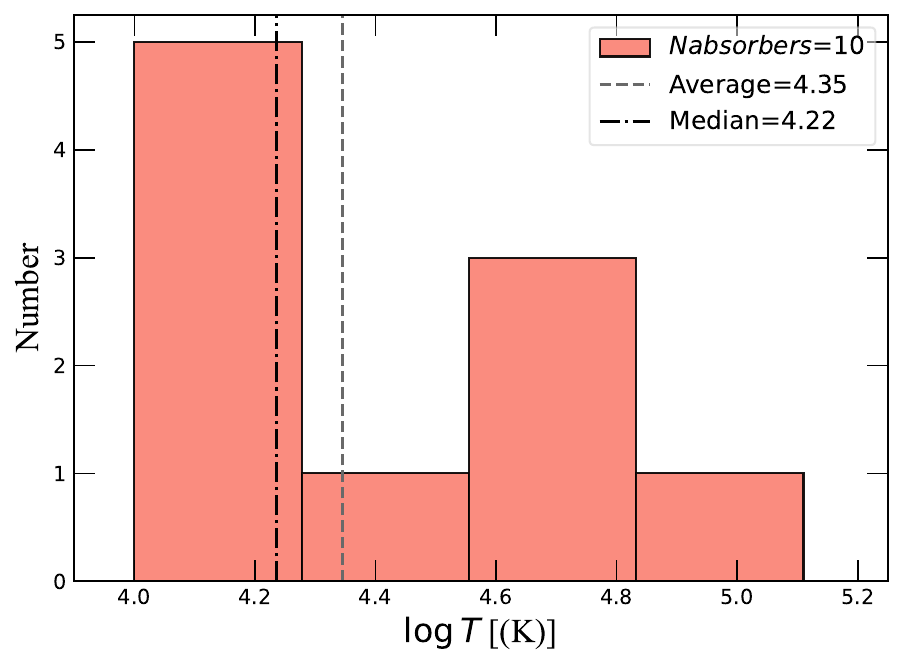}
    \includegraphics[width=0.48\linewidth]{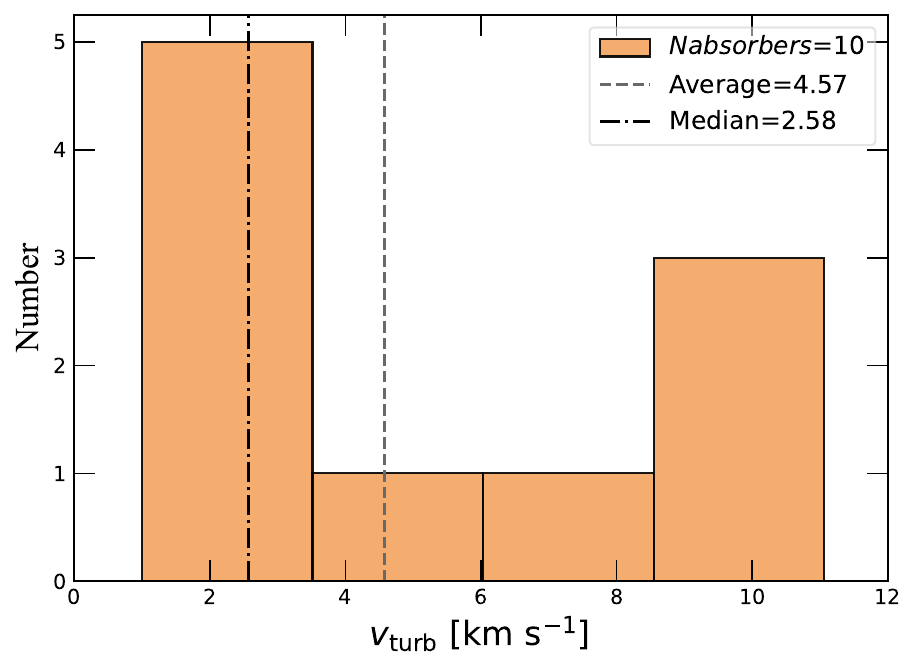}
    \includegraphics[width=0.48\linewidth]{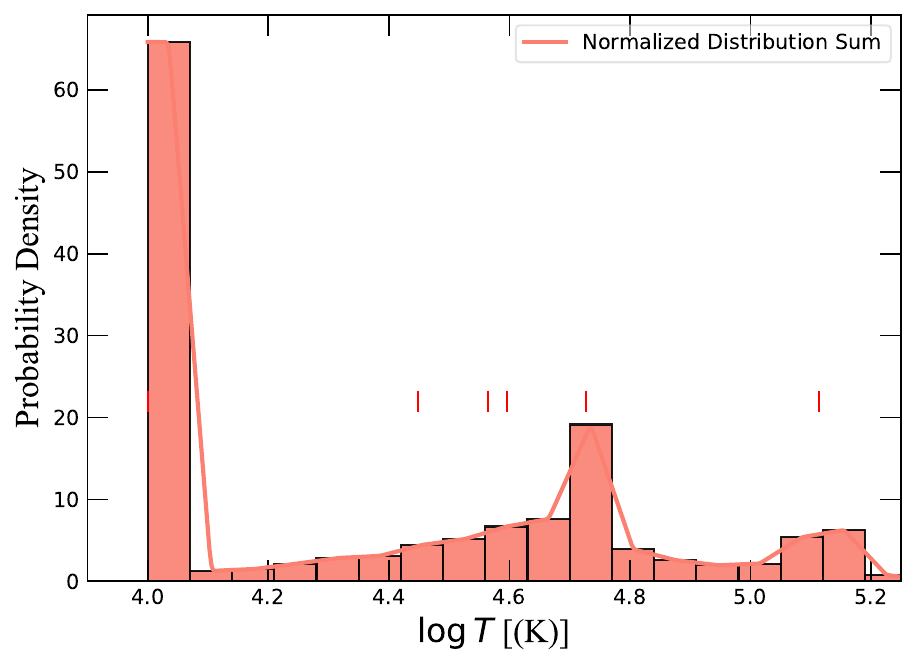}
    \includegraphics[width=0.48\linewidth]{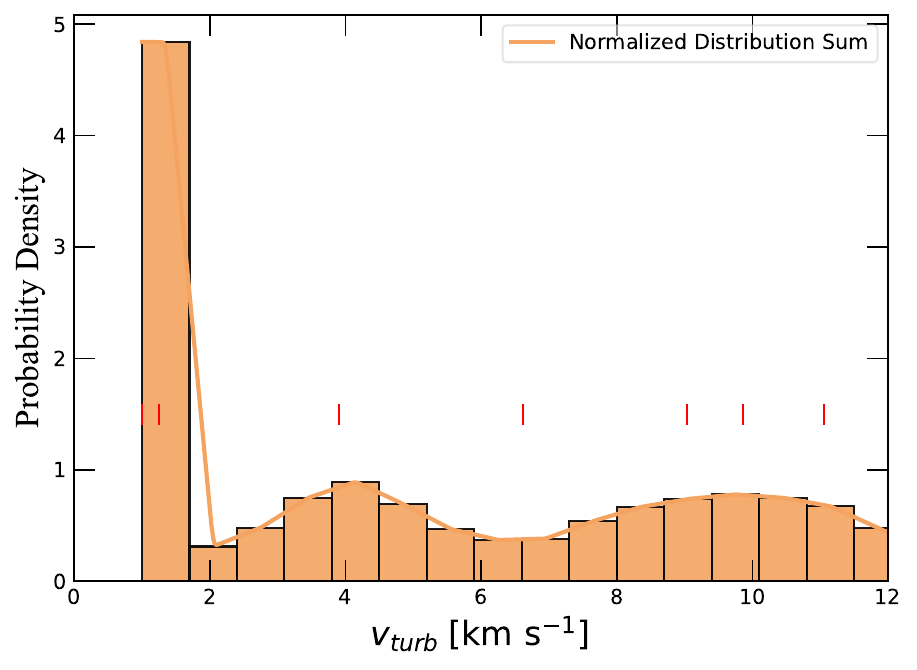}
    \caption{\textnormal{Distributions of the best-fit $\log{T}$ (\textit{left}) and \vturb\ (\textit{right}) for all thermally tied absorbers associated with galaxies HU1 and BX426b.} \textnormal{The \textit{top row} shows the adopted best \T\ and \vturb\ as histograms which are the median \T\ and \vturb\ of each Monte Carlo simulation run discussed in Appendix \ref{sec:apdx_vperrors}. The grey vertical dashed lines shows the average of the distributions, the black dot-dashed vertical lines show the medians, these are the same for each column. The $1\sigma$ scatter for each distribution is $\sigma(\logTK)=0.38$ and $\sigma(\vturb)=3.96~\kms$.} \textnormal{The \textit{bottom row} shows the probability density function which is the sum of the normalized parameter distributions (area equal to 1) for each individual component to visualize the uncertainty of each component combined. The red vertical lines denotes the location of the adopted \T and \vturb\ (median of the distributions). The typical error (median) for each component is fairly small; the typical error in \logT\ is $\sigma(\logTK)=0.21\pm0.11)$ while the typical error in \vturb\ is $\sigma(\vturb)=2.23\pm2.18\kms$}. We can see that the majority of the absorbers (6/10) are cool \textnormal{at temperatures less than $\logTK \leq 4.4$ while the other absorbers are warm-hot with temperatures ranging between $4.6 \leq \logTK\leq 5.1$ suggesting that this gas will cool rapidly. All of the absorbers had small turbulent velocities less than $\vturb \leq 11.0~\kms$ which suggests that the gas has subsonic internal velocity dispersion.}}
    \label{fig:T_bturb_hist}
\end{figure*}

Figure \ref{fig:T_bturb_hist} shows the \logTK\ and turbulent velocity \vturb\ \textnormal{distributions} of all the thermally tied components for HU1 and BX426b. \textnormal{The top panels show histograms of the best-fit \T\ and \vturb\ while the bottom panels visualize the typical error in each component by plotting the probability density function (sum of the normalized individual parameter distributions of all components) with markers that show the adopted best \T\ and \vturb\ for each component (median from the distributions).} 

The left \textnormal{column panels show the temperature distributions where} we see that the typical absorber is best fit with cool $\logTK \sim 4$ gas, consistent with other low-\nhi\ absorbers where the dominant mode of heating is from photoionization from the UV background at $z\sim2.245$ \citep[e.g.,][]{rudie+2012b}. \textnormal{The bottom panel shows that the errors are generally small for each component with the typical error (median) in \logT\ being $\sigma(\logTK)=0.21\pm0.11)$ (average \logT\ error is $\overline{\sigma(\logTK)}=0.23$).} Interestingly, 4/10 of the absorbers showed intermediate temperature gas $\logTK =4.6-5.1$ which is near the peak of \textnormal{H, C, and O's} cooling curves \citep[e.g.,][]{schure+2009}. \textnormal{The precise cooling time of the gas depends on density, metallicity, equilibrium assumptions, redshift, etc. but assuming a density of $n_{\rm H} = 10^{-4} ~ \cmthree$, collisional ionization equilibrium, and metallicity between $Z=0.3-1~ Z_\odot$, the hottest gas ($\logTK=5.1$) will cool between $t_{\rm cool}=21-33~{\rm Myr}$ depending on the cooling function used \citep[e.g.,][]{benson+2002,wiersma+2009}. The free fall time of the gas is $t_{ff} \approx 5~{\rm Gyr}$ which implies that $t_{\rm cool}/t_{\rm free} < 6\times10^{-3}$. The dynamical time of the galaxies (based on their measured sizes and nebular linewidths in Section \ref{sec:ISM}) are $t_{\rm dyn}({\rm HU1})=28~\Myr$ and $t_{\rm dyn}({\rm BX426b})=38~\Myr$ such that for both galaxies $t_{\rm dyn}/t_{\rm cool}\sim 1$. This all suggests that the gas will cool rapidly.}

The right \textnormal{column panels show the turbulent velocity distributions where wee see that the majority of the absorbers (6/10) are best fit with low velocities $\vturb \sim 1-4~\kms$. The bottom panel shows that the errors are generally small to moderate for each component with the typical error being $\sigma(\vturb=2.23\pm2.18\kms)$ (average \vturb\ error is $2.94~\kms$).} These small velocities are always paired with high $T$ except for \textnormal{one component whose linewidth was narrow enough ($b_d=5.19\pm1.84~\kms$) that it} preferred low $T$ \textit{and} low $\vturb$. We see the opposite behavior with components with large $\vturb$ in that they preferred small $\logTK \sim 4$; these components were associated with the highest \lognhi\ components. \textnormal{Given the small turbulent velocities that we measure, the internal velocity dispersion of the gas is likely subsonic.}

\begin{figure}
    \centering    
    \includegraphics[width=1.0\linewidth]{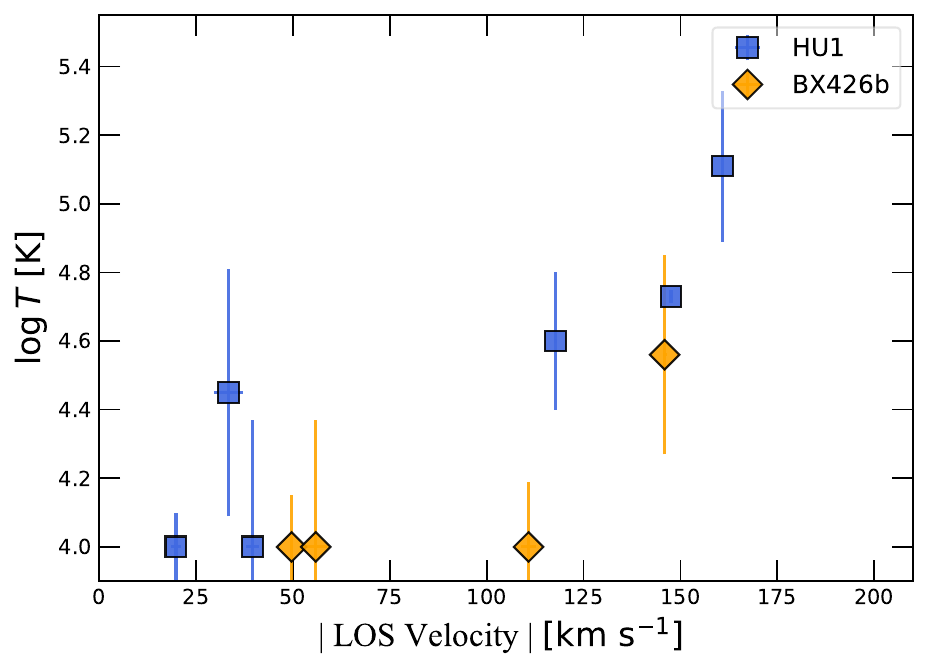}
    \includegraphics[width=1.0\linewidth]{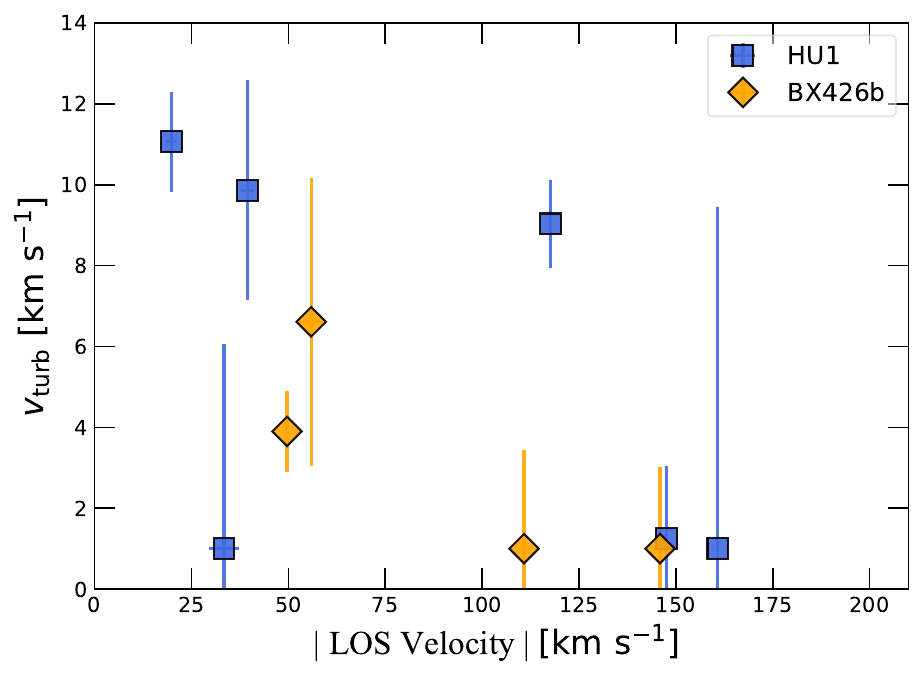}
    \caption{\textnormal{Best-fit $\log{T}$ and \vturb\ as a function of absolute line-of-sight velocity for all thermally tied absorbers.} The error bars show the 1$\sigma$ spread of best-fit values from the simulation \textnormal{posteriors} discussed in Appendix \ref{sec:apdx_vperrors}\@. \textnormal{All of the absorbers are tied with \CIV\ and at least \HI, \SiIII, \SiIV, or \OVI.} There is a trend with large absolute LOS velocity \textnormal{and} high temperature.}
    \label{fig:Tlosv_vturblosv}
\end{figure}

\textnormal{Figure \ref{fig:Tlosv_vturblosv} shows \logT\ and \vturb\ as a function of absolute line-of-sight velocity $|\vlos|$ for all thermally tied absorbers. The top panel shows \logT\ as a function of \vlos\ which hints at a trend between increasing $|\vlos| > 0 ~ \kms$ and increasing temperature.} This trend can be explained with a galactic outflow origin where the components are fast moving hot gas that was expelled/ejected from the galaxy in the past. Indeed, we showed in previous sections (Section \ref{sec:ISM_nebular} and \ref{sec:ISM_morphology_stellarpop}) that the galaxies are moderately star-forming based on their sSFR and could be driving outflows based on their \SFRSD. This combined with the small projected distances of the galaxies from the QSOs suggests that the CGM gas could be affected by galactic feedback.

The bottom panel of Figure \ref{fig:Tlosv_vturblosv} shows \vturb\ as a function of \vlos. A strong trend is not present but it appears that components with the highest \vturb\ are located at small $\vlos=0~\kms$, have low \T, and have the largest column densities in their halos. These high \vturb\ components are not easily explained by galactic outflows because their high turbulence should be paired with high temperatures, which we do not detect \citep{schmidt+2021}. Instead, they could be co-spatial with the high-\HI\ absorbers which would preferentially trace denser, cooler gas, where the bulk internal motions of the gas dominates the broadening.

An important note here is that for the \logT\ measurements near $\vlos=0~\kms$, only C and O were available to place thermal constraints (because \HI\ could not be tied to any individual metal component due to saturation). These elements are similar in mass with a ratio of $\sqrt{m_{\rm{O}}/m_{\rm{C}}}=1.15$ which is difficult to disentangle even with HIRES. \textnormal{Further, the high velocity components had the strongest \logT\ constraints because of the inclusion of \HI, which typically becomes non-saturated only at large \vlos. For example, the widths of the \HI\ components near $\vlos\sim0~\kms$ places upper limits on the temperature to be between $4.7 < \logTK < 5.3 \; (b_d = 31.9 \pm 1.1~\kms), b_d = 69.2 \pm 3.9~\kms)$, which is consistent with what we measured from \CIV\ and \OVI.} This illustrates the need to explore trends with \vlos\ with more halos using ions of sufficiently different masses to place stronger constraints on thermal properties of the absorbers at all \vlos\ e.g., C and Si with a mass ratio of $\sqrt{m_{\rm{Si}}/m_{\rm{C}}}=1.5$; \CIV\ and \SiIV\ absorbers are nearly ubiquitous in the $z\sim2.3$ CGM \citep{rudie+2017,rudie+2019,nunez+2024}.

\textnormal{We can compare the thermal and turbulent energy contributions inside the individual absorbers using the inferred \T\ and \vturb. Following the same derivation as \citet{rudie+2019}, the thermal energy contribution of the gas (per particle) is given by $E_{\rm Therm}=\frac{3}{2}kT$ for a monatomic gas where $k$ is the Boltzmann constant and $T$ is the temperature of the gas, the turbulent energy (per particle) is given by $E_{\rm turb}=\frac{1}{2}\mu m_p\vturb^2$ where $\mu=0.6$ for the mean molecular weight, $m_p$ is the mass of the proton, and \vturb\ is the turbulent velocity of the gas. These two energies can be connected via the sound speed $c_s=\frac{5kT}{3\mu m_p}$ (for a monatomic gas) and Mach number $M=\vturb/c_s$. Therefore, the ratio between the thermal energy contribution and total contribution is given by}

\begin{equation}
\frac{E_{\rm Therm}}{E_{\rm Therm}+E_{\rm turb}}=\frac{1}{1+\frac{5}{9}M^2}
\end{equation}

\begin{figure}
    \centering
    \includegraphics[width=1.0\linewidth]{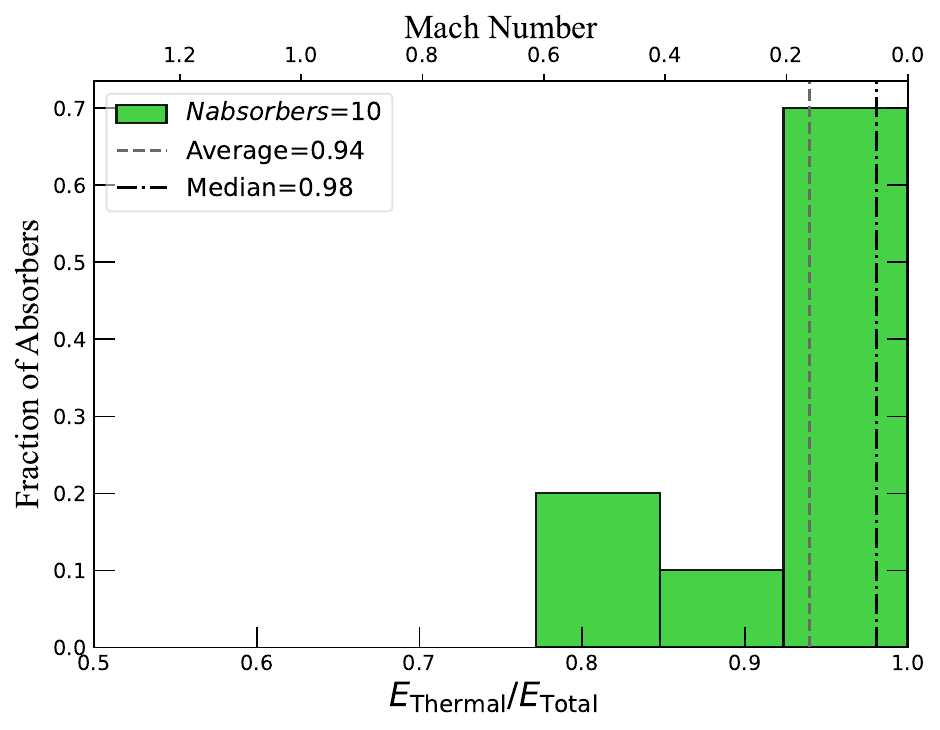}
    \includegraphics[width=1.0\linewidth]{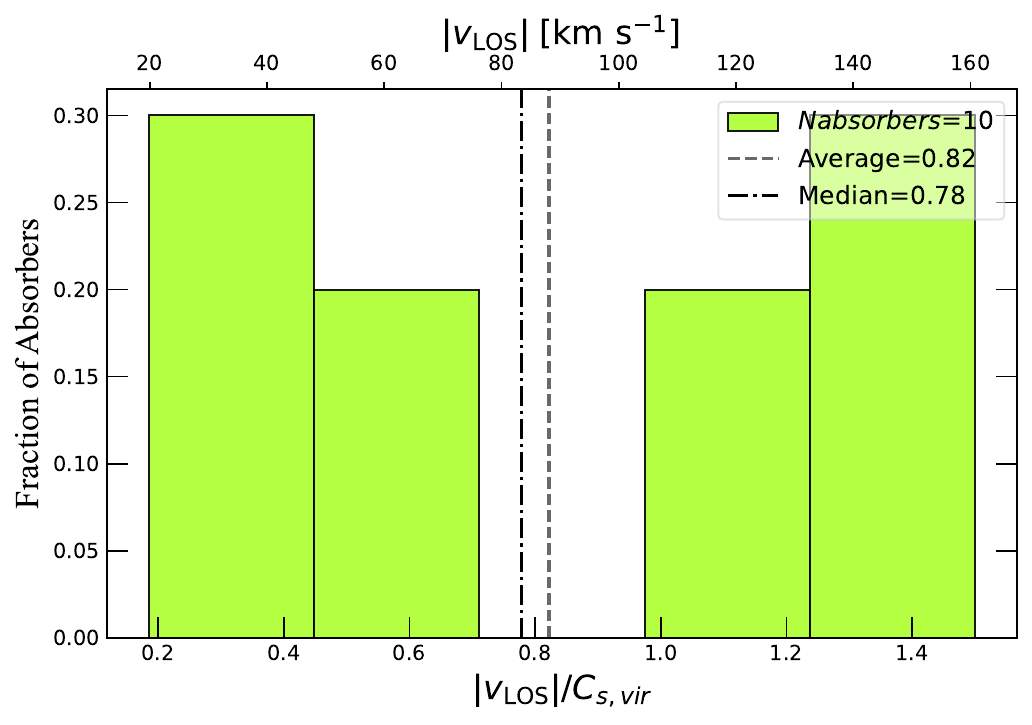}
    \caption{\textnormal{Energy and mach properties of the thermally tied absorbers. The grey dashed lines show the average of the distributions while the vertical black dash-dotted lines show the median. \textit{Top panel:} Thermal energy contribution compared to the total energy (thermal + turbulent) of all the thermally tied absorbers. The top axis shows the equivalent Mach number. The $1\sigma$ spread of the distribution is $\sigma(E_T/E_{TOT})=0.08$. All of the absorbers are dominated by thermal energy and all of the absorbers show subsonic (internal) turbulence. \textit{Bottom panel:} Histogram of line-of-sight velocity (bulk motion velocity) mach number. We assume that the absorbers are embedded in a virialized halo with a volume-filling medium that is similar in composition to the absorbers (same mean molecular weight) but with a higher temperature $T \sim \Tvir$ ($T_{vir}=10^{5.7} ~ \K$); the corresponding sound speed is $C_{S,vir}=107~\kms$. The top axis shows the equivalent line-of-sight velocity. the $1\sigma$ spread of the distribution is $\sigma(\vlos/C_{S,Vir})=0.47$. We can see that half of the absorbers have bulk velocities that are supersonic; these absorbers also have the highest temperatures.}}
    \label{fig:Etot}
\end{figure}

\textnormal{The top panel of Figure \ref{fig:Etot} shows the contribution of the internal thermal energy compared to the total internal energy (thermal and turbulent) budget for all thermally tied absorbers. We can see that thermal energy dominates the total energy budget for each absorber ($E_{\rm Th}/E_{\rm TOT} \geq 0.8$). Alternatively, the turbulent energy is subdominant and never exceeds $E_{\rm turb}\leq0.25 E_{\rm Therm}$; the typical $E_{Th}/E_{\rm TOT}$ is very close to 1. The top axis of the plot shows the equivalent mach number for each absorber which clearly shows that all of the absorbers have internal turbulent velocities that are subsonic. In other words, the \T\ and \vturb\ of each absorber analyzed in this sample has total internal energy that is dominated by thermal broadening ($E_{\rm Therm}>>E_{\rm turb}$) with turbulent (internal) velocities that are always subsonic ($M<1$). We compare these results to other studies in Section \ref{sec:discussion_literature}.}

\textnormal{The bottom panel of Figure \ref{fig:Etot} shows the distribution of line-of-sight velocity mach number, $M_{\rm LOS} = \vlos / C_{S,Vir}$ where $C_{S,Vir}$ is the sound speed at the virial temperature of the halo. If we assume that the absorbers are embedded in a virialized halo with a volume filling medium that is similar in composition to the absorbers (same mean molecular weight $\mu=0.6$) at the halo virial temperature $T\sim\Tvir$ where \Tvir\ is approximately $T\approx10^{5.7}~\K$ (see Section \ref{sec:discussion_halo}), then the corresponding sound speed is $C_{S,Vir}=107~\kms$. Under this assumption, we can see that half of the absorbers have bulk motions that are supersonic. The supersonic absorbers have the fastest bulk velocities and the highest temperatures in the sample (see top panel of Figure \ref{fig:Tlosv_vturblosv}) which may point to shocks as one of the heating mechanisms of the gas \citep[e.g.,][]{faucher-giguere+2023}. However, we note that there is no observational smoking gun evidence of a single, hot ($T\sim \Tvir$) volume filling phase in the CGM thus far \citep[e.g.,][]{tumlinson+2017}, \vlos\ is only one vector of the 3D velocity of the gas, and the virial temperature that we adopt is an upper limit because it is based on upper limits on the galaxy's stellar masses which all implies that more of the absorbers may actually be supersonic. Importantly, this implies that the supersonic gas is robust against these possibilities. Regardless, these findings point to a picture where the fastest-moving CGM gas was driven outwards with a hot outflow and is now moving supersonically through the halo. We compare these results to other studies in Section \ref{sec:discussion_literature}.}

\section{Novel and Preliminary Insights on the Low-Mass CGM at $z\sim2.3$} \label{sec:discussion}
In this Section we couple our ISM and CGM analyses to take the first steps towards constraining the galaxy-scale baryon cycle of low-mass star-forming galaxies at $z\sim2.3$. We then compare our findings to two studies that investigated halo gas within $\sim \rvir$ for massive $z\sim2.3$ galaxies and $z\sim0.3$ dwarf galaxies. 

\subsection{Connecting Inferred DM Halo Properties with the Halo Gas} \label{sec:discussion_halo}
We want to confirm if the QSO is within \rvir\ of each galaxy. We use the dynamical masses of HU1 and BX426b as upper limits on their stellar masses to calculate their halo masses (see Section \ref{sec:data_virial}). We find that BX426b has a halo mass of $\log{(M_h/\msun)} = 11.51^{+0.01}_{-0.04}$. For HU1 we calculate a range of halo masses using the dynamical masses of the different emission knots which give $\log{(M_h/\msun)}_{\rm Knot~A,B} < 11.13$ and $\log{(M_h/\msun)}_{\rm Knot~C} = 11.17^{+0.03}_{-0.04}$. These halos masses differ by more than 0.5 dex of the typical halo masses determined for more massive star-forming galaxies in KBSS with $M_{\rm halo}\sim10^{12}~\msun$ and stellar masses of $\logMmsun\sim9.8$ \citep{trainor+2012}; this is also the typical median stellar mass of both the KBSS and the MOSDEF surveys \citep{runco+2022}.

Converting these halo masses to \rvir\ for BX426b gives $\rvir=67~\kpc$ and for HU1 of $\rvir({\rm HU1}) < 51~\kpc$. This implies that the QSO is within \rvir\ of BX426b where $(\Dtran/\rvir)_{\mathrm{BX426b}} = 0.75$. For HU1, the QSO is likely within \rvir\ depending on which emission knot is the host galaxy, where $(\Dtran/\rvir)_{\mathrm{HU1}} \geq 0.6$.

We calculate the virial temperature of the halos to be $\Tvir({\rm BX426b}) = 10^{5.8} ~ \rm{K}$, and $\Tvir({\rm HU1}) < 10^{5.6} ~ \rm{K}$. This is interesting because the hottest gas in each halo is $\log{(T_{\rm{max}}({\rm HU1})/\K)} \sim 5.1$ and $\log{(T_{\rm{max}}({\rm BX426b})/\K)} \sim 4.6$ which is between $\log{(T_{\rm photoionization}/\K)} \sim 4$ and \Tvir. The ions with which the temperatures were measured (H, C, Si, O) have cooling peaks near $\logTK \sim 5$ which makes the temperature difficult to explain because of the fast cooling time \textnormal{(see Section \ref{sec:CGM_Thermal}).} \textnormal{Therefore, this gas} would require some additional heating source besides the UV background and/or constant replenishment of the gas in this phase.

\subsection{Fraction of Unbound CGM Absorbers}
\label{sec:discussion_boundgas}
We calculate the escape velocity of the galaxies by assuming they are situated in NFW dark matter halos \citep{navarro+1997} at their projected distances from the QSO\textnormal{, and taking the square root of their potentials}. We adopt halo concentrations of c(BX426b)=3.9 and c(HU1)=4.1 which was inferred from their halo masses input into the scaling relations derived by \citet{duffy+2008}. \textnormal{Adopting these concentration parameters to the NFW halo, we calculate the escape velocities to be $v_{\rm esc}(\Dtran=50~\kpc) = 210~\kms$ for BX426b, and $v_{\rm esc}(\Dtran=31~\kpc) = 199~\kms$ for HU1.}

\begin{figure}[tp]
    \centering   
    \includegraphics[width=0.98\linewidth]{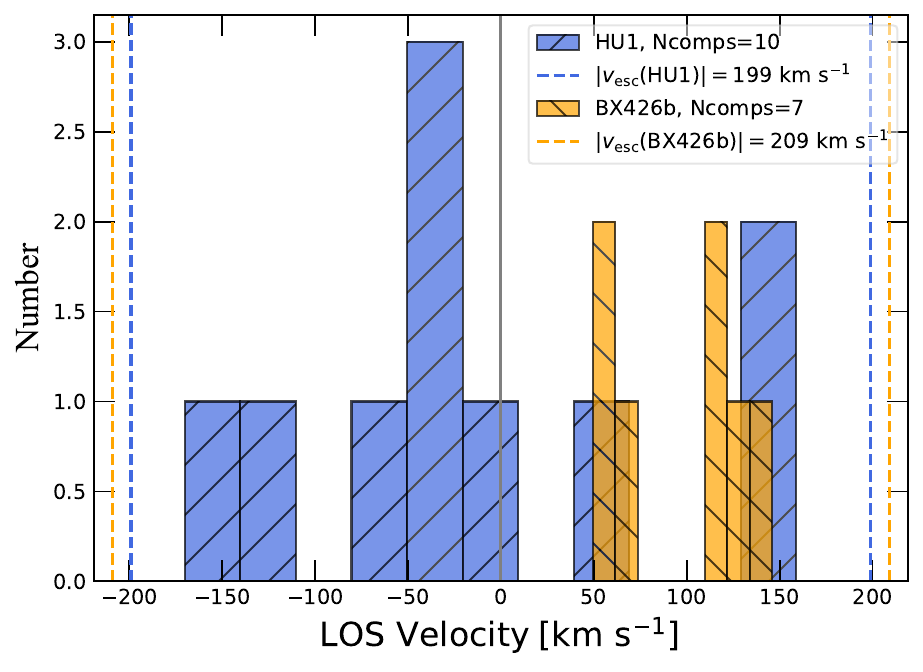}
    \caption{\textnormal{Histogram of the line-of-sight velocity for each CGM metal absorber analyzed in this work.} \textit{Blue (``/'' hatching)} shows galaxy HU1 and \textit{\textnormal{gold} (``\textbackslash''\ hatching)} shows BX426b. The escape velocity of each galaxy at their projected distance is shown as a dashed colored line. \textnormal{The errors of each absorbers are much smaller than the symbol depicting their line of sight velocity.} We can see that HU1 (blue) has one component that is unbound. 
    }
    \label{fig:vesc_hist} 
\end{figure}

Figure \ref{fig:vesc_hist} shows the distribution of line-of-sight velocities for each CGM metal absorber analyzed in this work. The colors correspond to absorbers in HU1's halo (blue) and BX426b's halo (\textnormal{gold}). For both galaxies we can see that none of the absorbers cross their halos $v_{\rm esc}$ line which shows that none of its CGM gas is \textit{unambiguously} unbound.

\textnormal{There are some components that are near to the escape velocity e.g., $|v_{\rm{HU1}}| = 170~\kms$.} It is possible, if not likely, that this high-velocity gas has a true velocity that is larger than the component projected along the line of sight that we measure. The LOS velocity is a lower limit of the three dimensional velocity vector. Additionally, the escape velocity that we quote is an upper limit because the transverse distance that we measure is a lower limit \textnormal{of the three dimensional distance to the CGM gas meaning the gas may be in front of or behind the galaxy} which would increase $r$ \textnormal{and decrease $v_{\rm esc}$}.

Given these arguments, it might be the case that the absorbers with the largest velocities $|v_{\rm{LOS}}|\sim|v_{\rm{esc}}|$ are unbound. But, we have no \textit{direct} evidence for this scenario since we can measure only \vlos. 

\textnormal{We acknowledge that it is likely that the stellar masses that we adopted for each galaxy (based on their dynamical masses) are overestimates. This would affect the escape velocity calculation and the resultant unbound gas escape fraction. To investigate this, we recalculate the stellar mass of each galaxy assuming the typical stellar mass to dynamical mass ratio ($\Mstar/\Mdyn$) of $z\sim2.3$ star-forming galaxies, which ranges between 0.27-0.32 \citep{maseda+2013,price+2020}; however, systems can show $\Mstar/\Mdyn < 0.1$ or $\Mstar/\Mdyn \sim 1$ \citep[e.g.,][]{erb+2006b}. If we adopt $\Mstar/\Mdyn = 0.3$ then the stellar masses of the galaxies range between $\logMstar/\msun$=8.32-8.55. This yields a new escape velocity for BX426b of $v_{\rm esc}=167~\kms$ which shows that the CGM absorbers are still bound to the galaxy since the fastest moving absorber has a slower velocity of $|v_{\rm max}|=146~\kms$ (Figures \ref{fig:stackplot_BX426b} and \ref{fig:Tlosv_vturblosv}). Recalculating the escape velocity for HU1 yields $v_{\rm esc}=161~\kms$ suggesting that the fastest moving component is unbound since $|v_{\rm max}|=170~\kms$. We caution that this result depends strongly on the adopted $\Mstar/\Mdyn$ ratio which has a significant amount of scatter that depends on e.g., stellar mass to gas mass fraction, specific star formation rate \citep[e.g.,][]{price+2016}. Therefore, we take this result as tentative, especially considering that adopting $\Mstar/\Mdyn \geq 0.38$ would result in the gas being bound. In other words, we currently have no strong definitive evidence for unbound gas in the halo of either galaxy. Deeper photometry in multiple bands would improve  confidence in this result.}

The \textnormal{conclusion} that neither galaxy has \textnormal{\textit{unambiguously} unbound gas in its CGM} is interesting because the typical range of wind velocities expected from stellar feedback driven galactic outflows are a few hundred \kms\ but can exceed $v_{\rm SN,Max} \gtrsim 1,000~\kms$ \citep{thompson+2024}. However, minimum speeds of $v_{\rm SN,Min} \gtrsim 100~\kms$ have been detected. Since $v_{\rm SN,Max}$ would well exceed $v_{\rm{esc}}$ for all halos at all redshifts, the typical and minimum wind velocities are a more meaningful comparison. The minimum velocity wind would not escape the gravitational potential of either galaxy in this sample but the typical wind velocity would. This motivates the need to investigate the wind properties \textit{and} unbound gas statistics of more low-mass $z\sim2.3$ galaxies to place constraints on e.g., mass propagation to the CGM in low mass galaxies vs high mass galaxies.

\subsection{Comparison with the Literature} \label{sec:discussion_literature}
We contextualize the CGM properties of HU1 and BX426b with three samples. \citet[][R19z2 hereafter]{rudie+2019} analyzed the inner CGM ($\Dtran<90~\kpc$) of eight $z\sim2, ~L_*$ star-forming galaxies. 
\citet[][M24z0 hereafter]{mishra+2024_CUBS9} recently analyzed one of the largest samples ($N=91$) of low redshift ($z\sim0.3$) low-mass/dwarf galaxies ($\logMmsun_{\mathrm{med}}\simeq8.3$) within a few \rvir\ of a QSO sightline ($\Dtran<300~\kpc$). This is a larger impact parameter range than the galaxies analyzed our work, so we restrict their sample to galaxies within projected distances of $\Dtran/\rvir \lesssim 1.25~\rvir \; $ of a QSO which includes measurements from \citep{johnson+2017}.
\citet[][N25z2 hereafter]{prusinski+2025} analyzed the largest sample of foreground galaxy-background galaxy pairs to map the structure of the CGM of $z\sim2$ star-forming galaxies ranging in stellar masses between $8\lesssim \logM \lesssim 11$ within 3 \rvir. We restrict our comparison to the galaxies within 1 \rvir.
\textnormal{\citet[][Q22z1 hereafter]{qu+2022_CUBSV} analyzed the thermodynamic properties $z\sim1$ massive galaxies in the CUBS survey \citep[Cosmic Ultraviolet Baryon Survey;][]{chen+2020_CUBS1}.}

\textit{The total \HI\ column densities (\nhi) for HU1 and BX426b are slightly greater than the median low-$z$ low-mass sample, but smaller than the median high-mass high-$z$ sample.} Specifically, the \HI\ in our work ranges between $\lognhicmtwo=15.1-15.7$. This overlaps with the total \HI\ column densities measured in the low mass low-$z$ galaxies in M24z0, which range from $\lognhicmtwo=13.8-17.6$ and is slightly larger than the median and average of their sample ($\lognhicmtwo_{\mathrm{M24z0,avg}}=14.8$, $\lognhicmtwo_{\mathrm{M24z0,med}}=14.5$). We calculated the total \nhi\ for the high-mass high-$z$ galaxies in R19z2 and find that it ranges between $\lognhicmtwo_{\mathrm{R19z2}}=15.7-20$ ($\lognhicmtwo_{\mathrm{R19z2,avg}}=17.0$, $\lognhicmtwo_{\mathrm{R19z2,med}}=16.4$); the median (and average) is larger than the range in our work.

\textit{The total metal column densities for HU1 and BX426b for intermediate ions is smaller than the high-mass high-$z$ sample whereas the high-ion column densities are similar to both the low mass low-$z$ sample and the high mass high-$z$ sample.} We found that both galaxies in our work have similar total \CIV\ column density of $\log{(N_{\CIV}/\cmtwo}) = 13.6$. The median \CIV\ column density of the high mass high-$z$ sample (excluding limits) is more than 0.7 dex larger at $\log{(N_{\CIV}/\cmtwo}) = 14.7$. Similarly, the halos in our work have similar total \OVI\ $\log{(N_{\OVI}/\cmtwo}) \sim 14.1$. Interestingly, the median of the low mass low-$z$ sample has a median \OVI\ column density of $\log{(N_{\OVI}/\cmtwo}) \sim 14.1$. The high mass high-$z$ sample only has 3 measurements of \OVI\ that range between $13.95 \leq \log{(N_{\OVI}/\cmtwo}) \leq 15.0 $ with a median of $\log{(N_{\OVI}/\cmtwo}) \sim 14.9$.

\textit{For HU1 and BX426b, we detect no low-ions (similar to low-mass low-$z$ galaxies), always detect intermediate-ions (similar to high-mass high-$z$ galaxies), and always detect high-ions (similar $z\sim0.3$ halos).} Specifically, we measure a high detection rate (fraction of galaxies that show a detection of an ion or phase) of unity for intermediate- (\CIV, \SiIII, \SiIV) and high-ions (\OVI), but a non-detection of low-ions (\SiII,\CII). These detection rates differ from both samples in that R19z2 found that high-mass high-$z$ galaxies show  moderate detection rate of low-ions of 38\%, high detection rate intermediate-ions of 88\%, and moderate detection rate of high-ions of 38\%. We note that the low detection rate of high ions is due to \lya\ and \lyb\ contamination from intervening \HI\ absorbers. M24z0 measured for the low-mass low-$z$ galaxies a low detection rate of low-ions 8\%, low detection rate of intermediate-ions of 21\%, but moderate detection rate of high-ions of 50\%.

\textit{HU1 and BX426b's CGM is more kinematically complex than low-$z$ low mass galaxies but more simple than high-mass high-$z$ galaxies.} We found that the kinematic component structure and velocity range of metals in HU1 and BX426b exhibit at least 6 unique \textnormal{metal} components (up to 10) spread over at least $|\Delta v|\sim150~\kms$ (up to $|\Delta v|\sim330~\kms$). This number-of-component and velocity range is smaller than but overlaps with R19z2's results that the $z\sim2.3$ halos exhibit at least 10 components (up to 25) spread over a velocity range of at least $|\Delta v|\sim 200~\kms$ (up to $|\Delta v|\sim 900~\kms$). Compared to M24z0, the range in this work is larger than but has overlap with their results which exhibit at most 4 components spread over at most $|\Delta v|\sim200~\kms$. In this case, the velocity spread of the absorption is the ideal quantity to compare against because resolving component structure is tied to both the resolving power and S/N of the QSO spectra which for M24z0 come from \textit{HST}/COS with $R\approx15,000$ and $S/N\approx12-31$ per resolution element \citep{chen+2020_CUBS1}, compared to this work (and R19z2) from HIRES with $R\approx45,000$ and $S/N\approx29-48$ per resolution element. 

\textit{The thermally tied absorbers in HU1 and BX426b's CGM have temperatures cooler than high-mass high-$z$ halos.} HU1 and BX426b's CGM components have temperatures that are consistent with heating from photoionization from the metagalactic background with a median temperature of $T_{\mathrm{med}}=10^4~\K$ (average temperature of $T_{\mathrm{med}}=10^{4.3}~\K$\textnormal{, with a $1\sigma$ scatter of $0.4~\K$)}, but span a wide range of temperatures: $4 \leq \logTK \leq 5.1$. Less than half of the absorbers (4/10) have intermediate temperature gas: $\logTK>4.6$. The measured turbulent velocities of all absorbers are fairly small with a median of $\bturb~_{\mathrm{,med}}=5.2~\kms$ (average of $\bturb~_{\mathrm{,avg}}=8.3~\kms$, \textnormal{with a $1\sigma$ scatter of $4~\kms$)} but also exhibit a wide range of velocities $\bturb=2.5-71~\kms$. Interestingly, all of these values are slightly lower than, but overall consistent with, high-mass high-$z$ galaxies which R19z2 measured a median temperature of $\logTK_{\rm med}=4.4$, a wide temperature range of $\logTK_{R19z2}=2.5-6.5$, a median turbulent velocity of $v_{\rm turb,med}=5.8~\kms$, and a wide turbulent velocity range of $\vturb = 1-70~\kms$. \textnormal{Q22z1 found that the temperature of the CGM absorbers in their massive $z\sim1$ sample had a typical (median) value of $\logTK=4.0$ with a scatter of 0.3 dex which is smaller than the high-$z$ sample, whereas their turbulent velocity has a typical value of $12~\kms$ with a scatter of $10~\kms$ which is larger than the high-$z$ sample.} M24z0 did not analyze the temperatures of the low-mass low-$z$ sample. \textnormal{There are other low-mass galaxies with detailed Voigt-profile based \T\ analyzed in the low-redshift ($z<1$) CUBS survey, and low-$z$ and $z\sim3$ MUSEQuBES survey. \citet[][CUBS III]{zahedy+2021_CUBS3} examined a Lyman Limit System (LLS; $17.2<\lognhicmtwo<19.0$) associated with two low-mass galaxies whose cool gas linewidths (probed by ${\rm Mg~II}$) implies temperatures of $(2-3)\times 10^4~\K$. \citet[][MUSEQuBES $z\sim0.5$]{johnson+2026} examined a LLS associated with a dwarf galaxy at $z=0.5723$ where they inferred temperatures of the low-ionization and high-ionization gas of $\logTK = 4.5\pm0.4$. \citet[][MUSEQuBES $z\sim3$]{banerjee+2025} examined a partial LLS (pLLS; $16.2<\lognhicmtwo<17.2$) associated with a LAE overdensity at $z\approx3.577$ finding that the temperature of gas as probed by \CIV\ and \SiIV\ was $\logTK = 4.7 \pm 0.2$. These temperatures ($4 < \logTK < 4.9$) are all consistent with but still cooler than those measured in this work ($4 < \logTK < 5.1$) which is interesting given their different redshifts ($z=0.5-3.5)$, different \HI\ column densities ($\lognhicmtwo=15.0-18.6$, consistency when using of low- and high-ions for each system, and that the gas often is well above the expectation for photoionization equilibrium $\logTK=4.0-4.3$. It will be important to investigate correlations of this trend with galaxy stellar mass, impact parameter, ion, LOS velocity, etc. }

\textnormal{\textit{The thermally tied absorbers in HU1 and BX426b's CGM absorbers have internal energies that are dominated by thermal broadening with subsonic internal velocities, similar to the high-mass high-$z$ sample.} All of HU1 and BX426b's CGM components have temperatures and turbulent velocities that imply that the thermal energy $E_{\rm Th}$ dominates the total energy $E_{\rm TOT}=E_{Th}+E_{\rm turb}$ with a high median ratio of $E_{Th}/E_{\rm TOT}=0.98^{+0.02}_{-0.08}$. This ratio is very similar to that found by R19z2 who measured a median of $E_{Th}/E_{\rm TOT}=0.97^{+0.01}_{-0.03}$. Q22z1 found that CGM absorbers in their massive $z\sim1$ sample showed a lower median thermal energy ratio of $E_{Th}/E_{\rm TOT}=0.82^{+0.41}_{-0.03}$. Unlike the higher $z$ samples which showed rather small $\vturb\sim4-6~\kms$, the $z\sim1$ sample showed a typical turbulent velocity of $\vturb=12~\kms$. Accordingly, the $z\sim1$ sample showed a much higher fraction of (internally) supersonic gas $>50\%$ compared to higher-$z$ $<30\%$. Taken at face value, this suggests that there may be a transition point between $z\sim1-2$ where turbulent broadening becomes more dominant.}

\textit{Neither \textnormal{HU1 nor BX426b possess} unambiguously unbound gas in their CGM.} \textnormal{We find an unbound gas fraction (number of galaxies that have at least one absorber with unbound gas) of $f_{\rm{vesc}}=0\%$.} The high-mass high-$z$ galaxies had a high escape fraction of $f_{\rm{vesc}}=71\%$ measured by R19z2. The escape fraction measured from low-$z$ low mass galaxies by M24z0 was low $f_{\rm{vesc}}=15\%$.

Our results corroborate recent work from N25z2 who found that the CGM properties of $z\sim2$ star-forming galaxies change strongly with stellar mass. Specifically, the equivalent width and line of sight velocity dispersion of \lya\ and \CIV\ decreases strongly with stellar mass at all transverse distances. Their lowest mass bin $\logMdynmsun = 9.0\pm0.5$ had both a lower normalization and steeper slope compared to the two higher mass bin they compare against ($\logMmsun=9.6\pm0.3, \; \logMmsun=10.2\pm0.4$. Taken together these findings point to a similar story where hydrogen and intermediate-ion equivalent width or column density, decreases with decreasing stellar mass. Similarly, line of sight velocity dispersion, or velocity range of absorption, decreases with decreasing stellar mass.

Our conclusions all point to a scenario where the low-mass $z\sim2.3$ CGM is low density, highly ionized, enriched with metals that have velocities that may not escape the galaxy gravitational potential, mostly has temperatures consistent with heating from the metagalactic UV background but also contains temperatures that require additional heating mechanisms (e.g., galactic winds) to sustain rapid cooling.

\section{Summary \& Conclusions} \label{sec:conclusions}
We have analyzed the first low-mass galaxies ($\mathrm{\Mstar \leq 10^{9}~M_\odot}$) in the KBSS-InCLOSE survey. The two galaxies, HU1 and BX426b, are in the same KBSS field towards QSO Q1623. Each galaxy is within a projected distance of $\Dtran \leq 50~\kpc$ ($\Dtran/\rvir <0.75$) of a luminous QSO (Figure \ref{fig:main_image}).

HU1 was discovered in a deep KCWI datacube (Figure \ref{fig:main_image}) while BX426b was discovered serendipitously on the same MOSFIRE slit as a more massive galaxy (Figure \ref{fig:pieces}). Analysis of their nebular spectra confirm that the galaxies are at a similar redshift $\zsys({\rm HU1})=2.2449, ~ \zsys({\rm BX426b})=2.2448$ (Sections \ref{sec:data_serendipity} and \ref{sec:ISM_nebular}) and star-forming based on detections of \lya\ and [O~III] emission $F({\rm O~III})<18\times10^{-18}~\fluxunits$ (Figures \ref{fig:spec_HU1} and \ref{fig:spec_BX426b}). 

HU1 is the most luminous \lya\ emitter in the InCLOSE sample thus far with $\log{(\lya/\lumunits)}\sim42$ and shows extended \lya\ emission out to a diameter of $d=30~\kpc$. Both HU1 and BX426b have small [O~III] line fluxes that are similar to that of faint $z \sim 2$ Lyman $\alpha$ Emitters (Table \ref{tab:spectra}) but are moderately star-forming based on limits placed on their specific star formation rates $\rm \sSFR<0.9~Gyr^{-1}$, and could drive outflows based on limits on their star-formation rate surface density $\SFRSD<0.4~\sfrsdunits$.

Joint analysis of the galaxies' nebular emission and morphology as seen in rest-optical \textit{HST} images confirm that the galaxies are low-mass based on their small projected sizes of $\rff \leq 0.9~\kpc$ (Figure \ref{fig:pieces}), and small dynamical masses of $\Mdyn({\rm BX426b})=10^{9}~\msun$ and $\Mdyn({\rm HU1}) \leq 10^{8.43}~\msun$ (Table \ref{tab:photometry}). 

The CGM of each galaxy as seen in HIRES QSO spectra shows weak \lya/\HI\ absorption ($\lognhicmtwo<15.6$), no low-ionization metal absorption (low-ions; i.e., \CII, \SiII), ubiquitous detections of intermediate-ions (\SiIII, \CIV, \SiIV), and ubiquitous detections of high-ions (\OVI) (Section \ref{sec:CGM}).

We performed Voigt-profile decomposition of the CGM absorbers showing that the \textnormal{metal-bearing} gas is composed of at least 7 kinematic components (up to 10) spread over a total velocity range of at least $150~\kms$ (and up to $|\Delta v|\sim330~\kms$) (Figure \ref{fig:stackplot_HU1} and \ref{fig:stackplot_BX426b}).

We inferred the thermal properties of a subset of CGM absorbers by decomposing their Doppler widths into temperature ($T$) and turbulent velocity \bturb. We found that the typical temperature of the absorbers is consistent with photoionization from the UV background ($\logTK_{\mathrm{med}}=4.0$; Figure \ref{fig:T_bturb_hist}) and that the typical turbulent velocity is small ($b_{\rm turb,med}=5.2~\kms$, Figure \ref{fig:T_bturb_hist}). Interestingly, 40\% of the absorbers with the largest velocities possess intermediate temperature gas that ranges between $4.6 \leq \logTK \leq 5.1$ that may have been ejected from a recent outflow (Figure \ref{fig:Tlosv_vturblosv}. Gas of this temperature would require an additional heating source or constant replenishment since this is near the peak of the ions' cooling curves \citep[H, C, O, Si; ][]{schure+2009,rudie+2019}. We also found gas nearest to the galaxy (in terms of line-of-sight velocity) line widths were dominated by turbulence (bottom of Figure \ref{fig:Tlosv_vturblosv}). \textnormal{We also showed that the internal energy of the CGM absorbers is dominated by thermal broadening, their internal turbulent velocities are subsonic, but their motions through the halo are likely supersonic.}

We coupled the galaxies' ISM properties with their CGM properties to gain preliminary insights into the galaxy-scale baryon cycle of low-mass galaxies at $z\sim2.3$. We found that each galaxy was within \rvir\ of the QSO $\Dtran<0.75~\rvir$ suggesting that we are likely analyzing inner CGM gas. The virial temperature of the galaxies, $\logTK>5.6$ is well above the intermediate temperature gas measured from the fastest moving absorbers ($\logTK<5.1$) showing that this gas is not long-lived. Neither galaxy showed evidence of unambiguously unbound CGM gas (Figure \ref{fig:vesc_hist}).

We contextualized our results with three studies that analyzed the halo properties of high mass $z\sim2.3$ galaxies (\citet{rudie+2019}), low-mass $z\sim0.3$ galaxies (\citet{mishra+2024_CUBS9}), and a sample of high- and low-mass galaxies at $z\sim2.3$ (\citet{prusinski+2025}). 
The CGM structure of the galaxies in our work is similar to high mass high-$z$ sample in its frequent detection of intermediate- and high-ions, and is similar to low mass low-$z$ galaxies in its non-frequent detection of low-ions.
The CGM kinematics of the galaxies in our work are more complex than the low-mass low-$z$ sample in terms of velocity range of metal absorption, but simpler than the high-mass high-$z$ sample in terms of number of components and velocity range.
The CGM thermal properties in our work are overall consistent with, but generally cooler than, those in the high-mass high-$z$ sample. Interestingly, both samples show intermediate temperature gas hinting that this may be a common feature of the $z\sim2$ CGM.
The turbulent velocities in this work are similar to the high mass high-$z$ sample in terms of total range and typical velocity.
\textnormal{Individual CGM absorber's internal energy is dominated by thermal broadening and their internal (turbulent) velocities are subsonic, similar to the high-mass high-$z$ sample.}
The fraction of \textnormal{\textit{unambiguously}} unbound gas detected in this work is more similar to the low-mass low-$z$ sample than the high-mass high-$z$ sample.

Taken together, the results suggest that the low mass $z\sim2.3$ CGM may be distinct from that of higher mass $z\sim2.3$ star-forming galaxies \textit{and} low mass $z\sim0.3$ star-forming dwarf galaxies. Its physical, kinematic, thermodynamic, and unbound gas properties favor values between the two populations. Even though this is a small sample, these results corroborate findings over the past decade that the CGM evolves strongly in redshift \citep[e.g.,][]{werk+2014,chen+2020_CUBS1,rudie+2012,rudie+2019,lofthouse+2020,muzahid+2020} and mass \citep[e.g.,][]{bordoloi+2014,johnson+2017,mishra+2024_CUBS9,prusinski+2025,banerjee+2026}. KBSS-InCLOSE will expand these preliminary results by exploring the largest sample of $z\sim2.3$ galaxy-QSO pairs within \rvir\ across a wide range of stellar masses ($\logMmsun = 8-10.5$) and galaxy properties, e.g., SFR, morphology, and ISM metallicity.

\section*{Acknowledgements}
We thank Zhiyuan Song for an insightful conversation that expanded and increased the quality of this work. We thank Ryan Cooke for explaining crucial points and limitations of the \textnormal{Monte Carlo simulation functionality} of ALIS which significantly improved the CGM modeling in this work. We thank Tommaso Treu, Allison Strom, and Brian Siana for insightful conversations that greatly improved the quality of this work. E.H.N. gratefully acknowledges support from the UC Presidents Office, Cal-Bridge Program, and Carnegie Observatories.

The authors wish to recognize and acknowledge the very significant cultural role and reverence that the summit of Maunakea has always had within the Native Hawaiian community. We are most fortunate to have the opportunity to conduct observations from this mountain.

\facility{Keck:II (KCWI)}
\facility{Keck:I (MOSFIRE)}
\facility{Keck:I (LRIS)}

\software{
ALIS \citep{ALIS},
Astropy \citep{astropy_2013,astropy_2018},
CubExtractor \citep{cantalupo+2019}
DS9 \citep{DS9_2000ascl.soft03002S},
IMEXAM \citep{IMEXAM_2022ascl.soft03004S},
MATPLOTLIB \citep{MATPLOTLIB_Hunter:2007},
MOSPEC \citep{strom+2017},
NUMPY \citep{numpy_harris2020array},
PHOTUTILS \citep{bradley_2022},
VoigtFit \citep{krogager+2017}
}

\bibliography{citations}{}
\bibliographystyle{aasjournal}

\appendix
\section{Quantifying Errors in Voigt Profile Fitting with ALIS} \label{sec:apdx_vperrors}
As discussed in Section \ref{sec:data_hiresvoigtprofile} there were \textnormal{metal-}absorption components whose best fit thermal and turbulent models were heavily dependent on initial starting parameters. In this section, \textnormal{we inspect the errors of the Voigt profile fits, then we quantify the effects of initial fitting parameters on the resultant best-fit \T, \vturb\ that are discussed} in Section \ref{sec:CGM_Thermal}.

\begin{figure*}
    \centering    
    \includegraphics[width=0.75\linewidth]{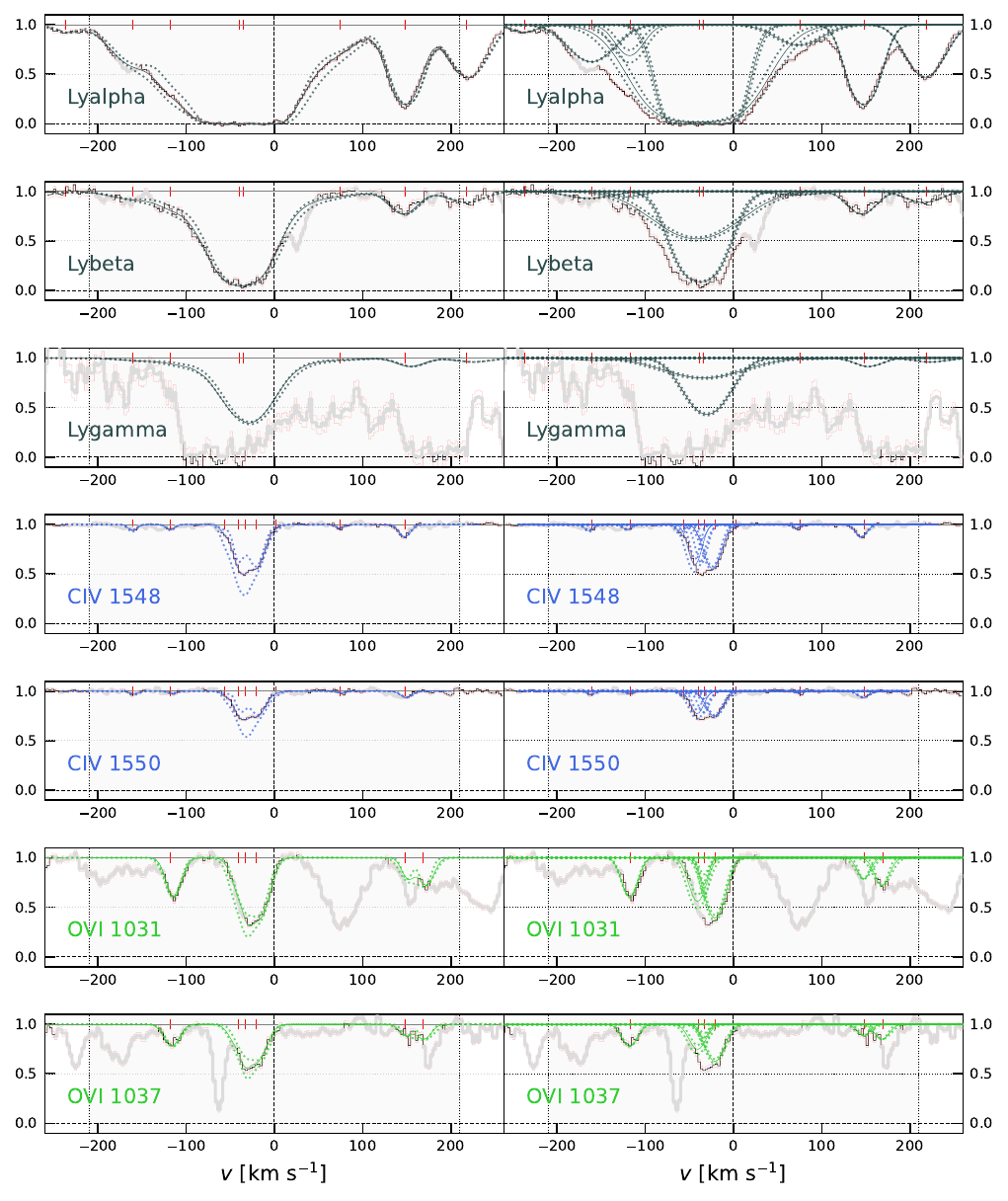}
    \caption{\textnormal{Ion stackplot showing the best-fit Voigt profile fits and associated errors for HU1. Same colors and lines as Figure \ref{fig:stackplot_HU1}. The dashed colored lines show the $\pm1\sigma~{log{N}}$ errors and $\pm1\sigma~b_d$ errors for each individual Voigt profile. The dashed lines above the main fit (solid colored lines) show $\log{N}+\sigma(\log{N})$ and $b_d+\sigma(b_d)$ where $\sigma$ is the $1\sigma$ error. Similarly, the bottom dashed line show $\log{N}-\sigma(\log{N})$ and $b_d-\sigma(b_d)$. We can see that the errors are small $\delta<0.3~\dex$ for the majority of the Voigt profiles.}}
    \label{fig:voigterror_HU1}
\end{figure*}

\begin{figure*}
    \centering    
    \includegraphics[width=0.75\linewidth]{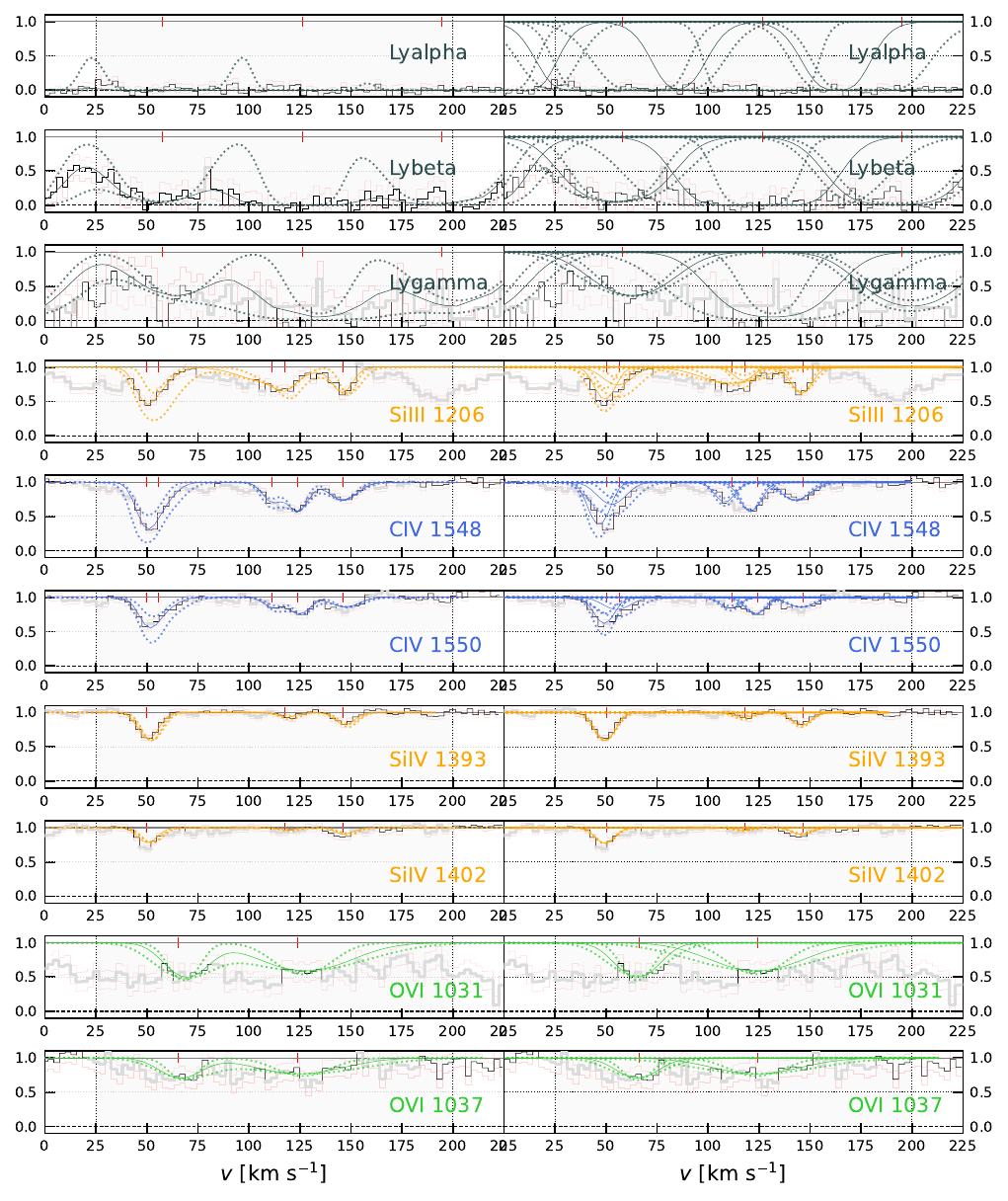}
    \caption{\textnormal{Ion stackplot showing the best-fit Voigt profile fits and associated errors for BX426b. Same colors and lines as Figure \ref{fig:voigterror_HU1}.}}
    \label{fig:voigterror_BX426b}
\end{figure*}

\textnormal{First, we visualize the Voigt profile parameter errors ($\log{N}, b_d$) in Figures \ref{fig:voigterror_HU1} and \ref{fig:voigterror_BX426b} by plotting the combined and individual best-fit Voigt profile models with their associated errors overplotted above and below the main fit as dashed lines. Specifically, the top dashed line shows $\log{N}+\delta(\log{N}), b_d+\delta(b_d)$ and the bottom dashed line shows $\log{N}-\delta(\log{N}), b_d-\delta(b_d)$. Note that we do not include the error in $z$ because it was well-determined for all components such that the error was much smaller than the resolution of the QSO spectra ($v_{res}=6.7~\kms$) with typical errors of $v\sim1.8~\kms$. The parameters have small errors of less than $\delta < 0.3 ~ \dex$ in $\log{N}$ and $b_d$ per component which can be seen when visually inspecting the fits. There are portions of the total fits (left column panels) where the errors appear to become larger than $> 0.3 ~ \dex$, near $v\sim0~\kms$ where the strongest constraints are present. This is an artifact of the plotting of the components where the errors are more closely plotted together leading to confusion between which error belongs to which component. When looking at the separated components (right panel column) we can see that the errors are in fact small for the individual components.}

\textnormal{Figure \ref{fig:voigterror_HU1} shows the associated errors for the CGM fitting of HU1. We can see that the majority of the components have small errors compared to the QSO error spectrum which translates to small errors in the Voigt profile parameters of less than $< 0.3 ~ \dex$ per component in $\log{N},b_d$; these errors directly translate to small errors in \T\ and/or \vturb. The \HI\ absorption is strongly constrained across the entire velocity range thanks to \lyb\ being only semi-saturated. The metal absorption has the smallest errors at high velocities where individual components are mostly isolated i.e., separated by large velocities which reduces covariance between the components - which is used to calculate the error in the Voigt profile fitting software (ALIS \citep{ALIS}). Consequently, the components with the largest metal errors are near $v\sim-25~\kms$ where the profiles are located more closely together leading to increased covariance. That said, the total fits on the left exaggerate the actual uncertainty in each component due to confusion between which error belongs to which component. This can be seen more clearly when visually inspecting the same velocity span in the right hand panel where the errors for each component appear more closer to their actual values.}

\textnormal{Figure \ref{fig:voigterror_BX426b} shows the errors associated with the CGM fitting of BX426b. We can see that the majority of the components have small errors across the spectrum for all ions. The errors for the \HI\ are larger than what was seen in HU1's halo due to \lyg\ being the only semi-saturated transition but with significant contamination which allows for a wider range of $\log{N},b_d$ that could achieve comparable goodness of fit. The metal absorption errors are the smallest for components at high velocity where the components are mostly isolated. The largest errors are found near $v\sim-50~\kms$ where the components are more closely spaced together. As with HU1, the total fit on the left exaggerates the errors slightly since there is confusion on which errors belong to which component. When inspecting the same velocity span in the right hand panel it can be seen more clearly that the errors are small; the largest errors are $< 0.3 ~ \dex$ in both $\log{N}$ and $b_d$.}

\textnormal{Next, we investigate the robustness of \T\ and \vturb\ against initial guesses. To do this, we ran} Monte Carlo simulations \textnormal{(without a Markov Chain)} to explore the uniqueness of the \textnormal{final best-fit \vturb\ and \T\ parameters}. We ran 1,000 simulations that used the best-fit parameters ($z,N,\vturb,T$; described in Section \ref{sec:CGM_Thermal}) as a starting point, then add random perturbations \textnormal{to each parameter ($z+z\Delta_{z,i},\log{N}+\log{N}\Delta_{N,i}, \vturb+\vturb\Delta_{v,i}, T+T\Delta_{T,i}$ where $\Delta_{X,i}$ is a random perturbation specific to each parameter $X$ and each iteration $i$}, and finally re-fit to the same goodness of fit criteria as \textnormal{before} (reduced $\chi^2 \sim1$). \textnormal{Note that each run is independent meaning that the best-fit solution of an iteration is not used as the starting point for the next iteration, so this is not a \textit{Markov Chain} Monte Carlo simulation. In other words, each iteration has the same starting point but a different random perturbation.}

\begin{figure*}
    \centering    
    \includegraphics[width=0.75\linewidth]{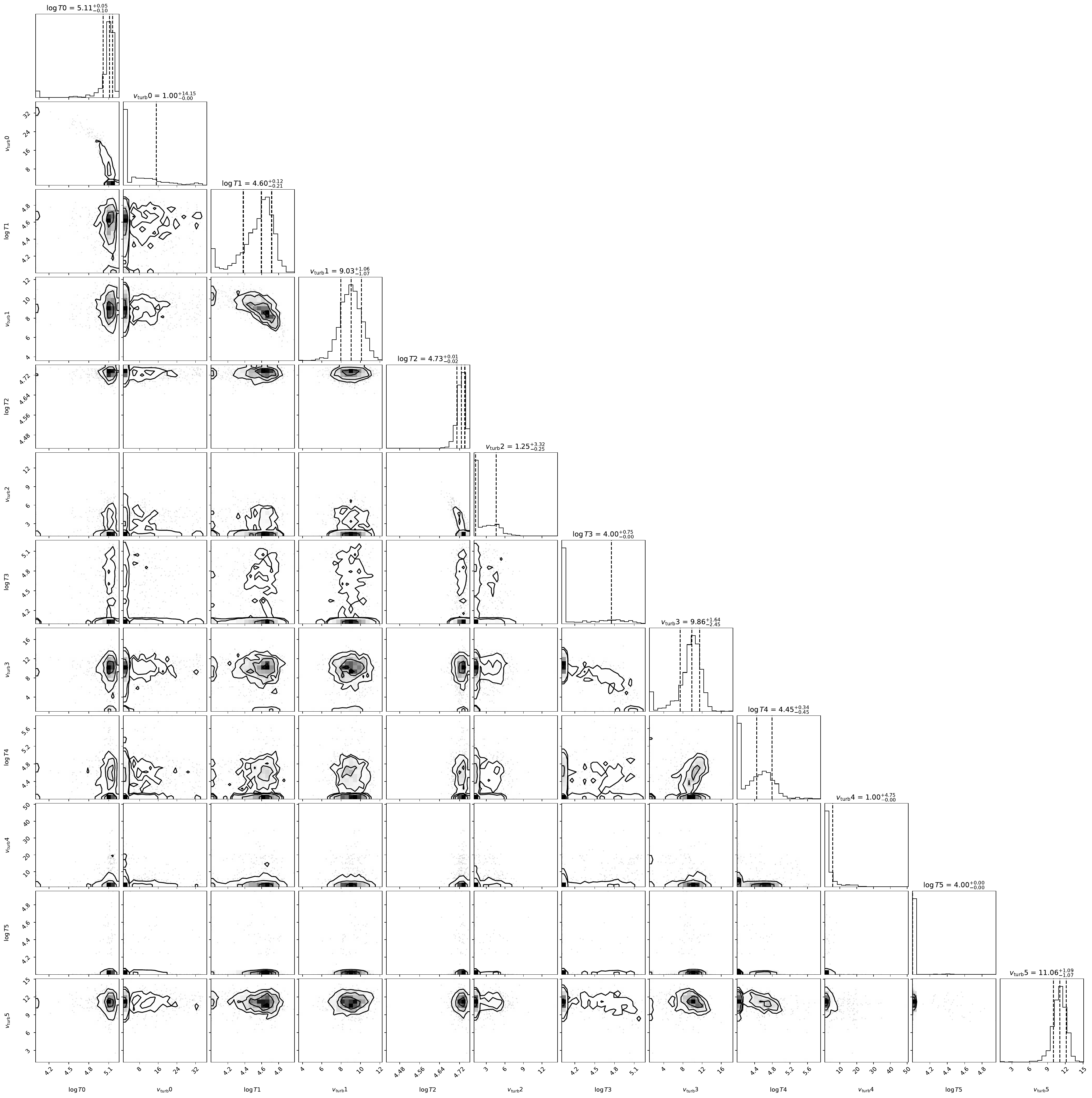}
    \caption{Corner plot summarizing the \textnormal{Monte Carlo} simulations performed on HU1's best-fit $\log{T}$ and $v_{\mathrm{turb}}$ for all \textnormal{thermally} tied absorbers. The simulation ran for 1,000 iterations and they \textnormal{all used the same starting but added random perturbations to each parameter for each iteration to describe the component structure shown in Figure \ref{fig:stackplot_HU1}.} We can see that all of the components have well-peaked distributions showing that the data are sufficient to \textnormal{determine the thermal and turbulent properties of the gas. The median and 1$\sigma$ spread in the distributions are adopted as the best-fit parameters and their errors in Section \ref{sec:CGM_Thermal}}.}
    \label{fig:corner_HU1}
\end{figure*}

\begin{figure*}
    \centering    
    \includegraphics[width=0.75\linewidth]{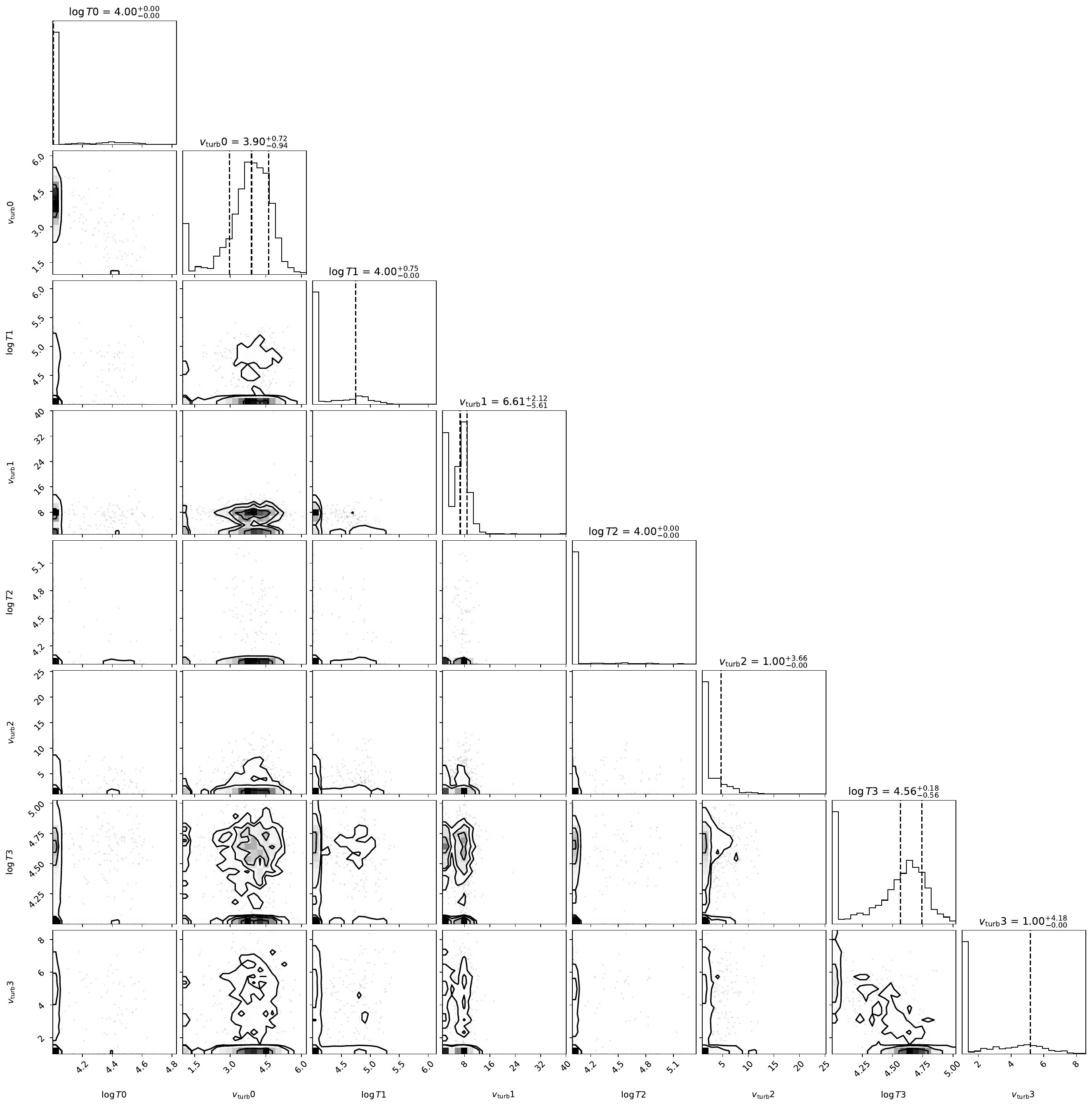}
    \caption{Corner plot summarizing the \textnormal{Monte Carlo} simulation performed on BX426b's best-\textnormal{fit} $\log{T}$ and $v_{\mathrm{turb}}$ for all \textnormal{thermally} tied absorbers using the same symbols and method described in Figure \ref{fig:corner_HU1}. We can see that only two of the components are well constrained by the data.}
    \label{fig:corner_BX426b}
\end{figure*}

\textnormal{We visualize the results of the Monte Carlo simulations in Figures \ref{fig:corner_HU1} and \ref{fig:corner_BX426b} which show corner plots of the \logT\ and \vturb\ distributions. We are primarily interested in how the distribution of \T\ and \vturb\ are related to one another per component and as such, the histograms on the top right of the corner plots, and the adjacent scatter plots (bottom of \T\ histogram and to the left of the \vturb\ histogram) provide the necessary insights for this goal. The other scatter plots provide insights on the correlations between the components which should be related but not strongly correlated; we see that this is indeed the case for both galaxies.}

In Figure \ref{fig:corner_HU1} we show \textnormal{the \T\ and \vturb} corner plot \textnormal{for} HU1's \textnormal{Monte Carlo} simulation \textnormal{that summarizes the distribution of parameters from the simulations}. \textnormal{For HU1's simulation we allowed the \T\ and \vturb\ to vary for five CGM} components that were tied with \CIV\ and \OVI\ and/or \HI, while the other components just varied $b_d$. When examining the e.g.,$\vturb,0-\log{T},0$ plots, we can see that all components have well peaked histograms that show clear preferences for whether the gas is \textnormal{dominated by thermal or turbulent broadening.} 

These results show that for \textnormal{the thermally-tied} components \textnormal{in HU1's halo}, the best-fit models are robust against the starting initial parameters, and that the best-fit parameters errors are not dominated by fitting systematics. In other words, \textnormal{the inferred thermal properties of HU1' CGM gas, \vturb, and \T, are well-constrained by the data.} We adopt the median of each \textnormal{distribution} as the best-fit \textnormal{\T\ and \vturb\ and adopt the 1$\sigma$ spread from the distribution as the error, then analyze and discuss} in Section \ref{sec:CGM_Thermal}.

In Figure \ref{fig:corner_BX426b} we show a corner plot of BX426b's \textnormal{Monte Carlo} simulation \textnormal{that was run using the same approach as HU1. The simulation allowed \T\ and \vturb\ to vary for four components that were tied between \CIV and \SiIV, and \SiIII\ while the other componets only $b_d$ was varied.} We can see that two out of four components have well peaked histograms suggesting that the data can reliably \textnormal{determine \vturb\ and \T}. However, two components do not show a clear preference. \textnormal{Specifically, component 3} shows a bi-modality in its best fit parameters but more often prefers a large temperature and small turbulent broadening; the errors in this component reflect this. Component 2 shows a clear preference of low \T\ and low \vturb\ suggesting that the widths are \textnormal{too} small enough \textnormal{for} the data \textnormal{to constrain whether thermal heating or turbulence dominates the broadening.}

\textnormal{These results show that half of the thermally-tied components in BX426b's halo are well-constrained in \vturb\ and \T\ by the data, whereas the other two components are not and have larger accompanying errors. We adopt the median of each distribution as the best-fit parameter, adopt the 1$\sigma$ spread from the distributions as the error, then analyze and discuss in Section \ref{sec:CGM_Thermal}.}

\section{Additional Emitters in the MOSFIRE Spectra} \label{sec:apdx_mosfireserendips}
\textnormal{In this appendix section we discuss} detected faint line emission that shared the same MOSFIRE slits as HU1 and BX426b that may or may not be physically connected to the galaxies. We discuss the consequences if \textnormal{the emission are} physically connected to the galaxies.

\begin{figure}
    \centering
    \includegraphics[width=1\linewidth]{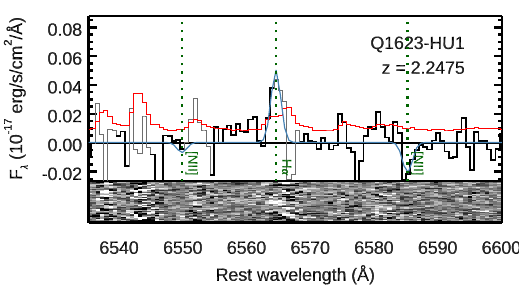}
    \caption{$K_{\mathrm{s}}$-band MOSFIRE spectrum with the same colors, symbols, and panels as Figure \ref{fig:spec_HU1}. There is a marginal detection of \Ha\ emission at a different redshift than what has been adopted \textnormal{from [O~III] ($\zneb=z_{\rm [OIII]} = 2.2449$)}.}
    \label{fig:spec_HU1ha}
\end{figure}

Figure \ref{fig:spec_HU1ha} shows HU1's K band spectra \textnormal{with a tentative} detection of \Ha\ emission at $z_{\Ha}=2.2475$ close to the correct slit position of HU1. \Ha\ falls very close to a sky line, which combined with the faintness of the emission, makes the characteristic upper and lower negative images from the ABAB mask nod difficult to detect (2D spectrogram in Figure \ref{fig:spec_HU1ha}). \Ha\ has \textnormal{a} measured linewidth \textnormal{that is} smaller than the instrumental resolution of MOSFIRE $\sigma < 35~\kms$ \textnormal{which} required us to freeze the linewidth to fit it with a 1D Gaussian. \textnormal{We measure} a slit-loss corrected flux of $F_{\Ha}=(5.32\pm1.41)\times10^{-18}~\fluxunits \; (\rm{S/N=3.7}$). This redshift corresponds to a velocity difference of \lya\ from systemic of $\Delta v_{\rm{\lya-sys}}=-74~\kms$. This suggest that \lya\ is blueshifted, \textnormal{which is not common \citep[e.g.,][]{trainor+2015}}. \textnormal{Most importantly}, this is a large velocity offset between $\mathrm{[O~III]}$ and \Ha\ of $|\Delta z_{\mathrm{[O~III]} - \Ha}|=231~\kms$ \textnormal{considering that both line should be probing the same gas}. We are unaware of any physical mechanism that would lead to such a velocity offset if one assumes that the gas is co-spatial. Further, if we adopt $\zsys=z_{\Ha}$, the velocity difference is not consistent with the CGM gas absorption velocity range. 

If one does assume that the emission does somehow arise from the same gas, the \lya\ to \Ha\ ratio is $F_{\lya}/F_{\Ha}=23.1/3.15=7.33$ which is close to Case B recombination ($I_{\lya}/I_{\Ha}=8.7$ assuming $T\sim10^4~K$ and $n_e\sim350~{\rm cm}^{-3}$; \citealt{osterbrock+1989}), leading to a very high \lya\ escape fraction $f_{{\mathrm{esc}, \lya}}=0.84$. This would be consistent with our finding that HU1 is similar to faint $z\sim2$ LAEs which are also characterized by such high $f_{esc}$ \citep{erb+2023}. However, our limit on \Hb\ leads to a non-physical \Ha\ to \Hb\ ratio of $F_(\Ha)/F(\Hb) < 1.6$ which is lower than the expected intrinsic ratio if there is no dust extinction ($I_{\Ha}/I_{\Hb}=2.79$, \textnormal{under the same temperature and density assumption}.

Taken together, the evidence favors a scenario where the \Ha\ emission, if real, is hosted by a different galaxy. We do not see any obvious continuum emitting objects in the F160W images along the slit (top right panel of Figure \ref{fig:main_image}) meaning the candidate galaxy galaxy has a magnitude greater than $m_{\rm F160W}\gtrsim27$. \textnormal{Alternatively, this could just be a sky line masquerading as a line emitter in which no physical insights can be gleamed from the emission. Deeper ground-based and/or space-based e.g., JWST, imaging and spectroscopic data would shed more light on these possibilities. Until then, we will not include the line emission in our analysis.}

\begin{figure}
    \centering
    \includegraphics[width=1\linewidth]{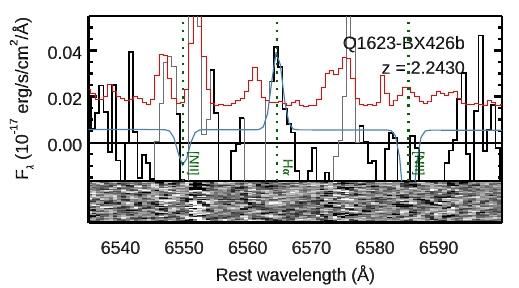}
    \caption{$K_{\mathrm{s}}$-band MOSFIRE spectrum of BX426b with the same colors, symbols, and panels as Figure \ref{fig:spec_HU1}. \textnormal{There is a marginal detection of \Ha\ emission at a different redshift than what has been adopted from [O~III]) ($\zneb=z_{\rm [OIII]} = 2.2448$).}}
    \label{fig:spec_BX426bHa}
\end{figure}

Figure \ref{fig:spec_BX426bHa} shows BX426b's K band spectrum \textnormal{with a tentative} detection of \Ha\ at a redshift of $z_{\Ha}=2.2430$\textnormal{, approximately at} the correct slit position of BX426b. \textnormal{Similar as HU1, this} redshift leads to a large velocity difference between [O~III] of $|\Delta v|=166~\kms$. \textnormal{As discussed previously, we are unaware of a physical mechanism that can explain such a large velocity difference.} We measure a flux of $F_{\Ha}=(3.56 \pm 1.38)\times10^{-18}~\fluxunits$ which gives a \textnormal{signal-to-noise} ratio of $\rm S/N\sim2.5$. The measured S/N is much lower than [O~III] at ${\rm SNR_{[O~III]}=9.1\sigma}$. This \textnormal{difference in S/N} can be seen \textnormal{more clearly} in the 2D spectrogram of Figure \ref{fig:spec_HU1ha} \textnormal{where there is a lack of} the two negative images that are characteristic of real detections in MOSFIRE spectra from the ABAB mask nodding. Lastly, if one adopts $\zsys=z_{\Ha}$ the velocity of the CGM absorption is not consistent with the strongest \HI\ and metal \textnormal{absorption} components. 

\textnormal{Altogether, the evidence suggests that if the line is real, the emission does not arise from the same gas as BX426b's [O~III] emission or it is an entirely different galaxy. An alternative explanation is that the line is not real and that we are seeing e.g., a hot pixel, masquerading as a line emitter. Deeper ground-based and/or space-based data would allow for the differentiation between these two possibilities. Until then, we will not include the line emission in our analysis.}

\section{Localizing the Emission Knots Responsible for HU1's Emission} \label{sec:apdx_emissionknots}
We observed HU1 (and BX426b) with three separate MOSFIRE slit masks, each with slightly different coordinates that prioritized each ``emission knot.'' Emission Knot B was included on all three masks (Q1623\_2022B, Q1623\_HU1mod\_2023B, Q1623sg1\_2024A) since it was located closest to the \lya\ \textnormal{centroid}. Specifically, Knot B was prioritized and centered in the first two masks (Q1623\_2022B, Q1623\_HU1mod\_2023B) which coincided with a \textnormal{marginal detection of HU1 ([O~III]$\lambda 5008$ emission at $z=2.2449$)}. It was at the very edge of the third mask (Q1623sg1\_2024A) where it had non-detection of the \textnormal{[O~III] emission}. 

Knot C was included in the last two masks (Q1623\_HU1mod\_2023B, Q1623sg1\_2024A) which coincided with \textnormal{a marginal detection of HU1 and} the detection of \textnormal{non-related} line emission at $z=2.2175$. It was not included in the first mask (Q1623\_2022B), and we do not see the \textnormal{unrelated emission at $z=2.2175$} in that spectrum. 

Knot A was only included on the first mask (Q1623\_2022B) since it was faintest knot and located furthest from the centroid. We detected no line emission that coincided with the inclusion of this knot on a slit.

\textnormal{Both Knot B and Knot C were included on a mask where HU1 was detected.} Therefore, it is appears that Emission Knot B or Knot C could be the host galaxy of the \lya\ emission seen in KCWI. \textnormal{Disentangling which emission knot is the actual host, or if both are the hosts, will require deeper ground-based and/or space-based rest-optical spectroscopy.}

\end{document}